\documentclass[10pt,journal,compsoc]{IEEEtran}

\ifCLASSOPTIONcompsoc
  \usepackage[nocompress]{cite}
\else
  \usepackage{cite}
\fi

\ifCLASSINFOpdf
\else
\fi

\usepackage{amsthm}
\usepackage{makecell}
\usepackage{amsmath,amsfonts}
\usepackage{graphicx}
\usepackage{subfigure}
\usepackage{multirow}
\usepackage{textcomp}
\usepackage{enumerate}
\usepackage{booktabs}
\usepackage{colortbl}
\usepackage{enumitem}
\usepackage{algpseudocode}
\usepackage{soul}
\usepackage{balance}

\usepackage[linesnumbered,ruled,vlined]{algorithm2e}

\SetCommentSty{mycommfont}
\SetKwInput{KwInput}{Input}                
\SetKwInput{KwOutput}{Output}              

\usepackage{hyperref} \hypersetup{colorlinks=true, linkcolor=black, filecolor=magenta, urlcolor=black, citecolor=black,hypertexnames=true}
\usepackage{layouts}

\usepackage{cleveref}
\crefformat{section}{\S\textcolor{purple}{#2#1#3}} 
\crefformat{subsection}{\S\textcolor{purple}{#2#1#3}}
\crefformat{subsubsection}{\S\textcolor{purple}{#2#1#3}}

\newlength\mylen

\usepackage{tcolorbox}

\newtheorem{Lemma*}{Lemma}

\definecolor{mygray}{gray}{.9}
\definecolor{mypink}{rgb}{.99,.91,.95}
\definecolor{mycyan}{cmyk}{.3,0,0,0}
\definecolor{coolblack}{rgb}{0.0, 0.18, 0.39}

\def\mathbi#1{\textbf{\em #1}}

\usepackage{tikz}

\makeatletter 
  \newcommand\figcaption{\def\@captype{figure}\caption} 
  \newcommand\tabcaption{\def\@captype{table}\caption} 
\makeatother

\newcommand{\todo}[1]{}
\renewcommand{\todo}[1]{{\color{red} TODO: {#1}}}

\graphicspath{{./image/},{./Figures/}}

\begin{document}
%
\title{EdgeDis: Enabling Fast, Economical, and Reliable Data Dissemination for Mobile Edge Computing}

\author{Bo~Li,~\IEEEmembership{Member,~IEEE,}
        Qiang~He,~\IEEEmembership{Senior Member,~IEEE,}
        Feifei~Chen,~\IEEEmembership{Member,~IEEE,}
        Lingjuan~Lyu,        Athman~Bouguettaya,~\IEEEmembership{Fellow,~IEEE,}
        and Yun~Yang,~\IEEEmembership{Senior Member,~IEEE}%

 \thanks{This research is funded by DP180100212, DP200102491, DP220101823, and LE220100078 grants from the Australian Research Council. The statements made herein are solely the responsibility of the authors.}       
\IEEEcompsocitemizethanks{\IEEEcompsocthanksitem Bo Li is with the College of Arts, Business, Law Education, and Information Technology, Victoria University, Melbourne, VIC 3122, Australia. \protect E-mail: bo.li@vu.edu.au
\IEEEcompsocthanksitem Qiang He is with the National Engineering Research Center for Big Data Technology and System, Services Computing Technology and System Lab, Cluster and Grid Computing Lab, School of Computer Science and Technology, Huazhong University of Science and Technology, Wuhan 430074, China, and the Department of Computing Technologies, Swinburne University of Technology, Melbourne, VIC 3122, Australia. \protect E-mail: hqiang@hust.edu.cn. Qiang is the corresponding author.
\IEEEcompsocthanksitem F. Chen is with the School of Information Technology, Deakin University, Geelong, Australia. E-mail: feifei.chen@deakin.edu.au. \protect
\IEEEcompsocthanksitem Lingjuan Lyu is with SONY AI Inc., Tokyo 108-0075, Japan. \protect E-mail: lingjuanlvsmile@gmail.com.
\IEEEcompsocthanksitem Athman Bouguettaya is with the School of Computer Science, University of Sydney, Camperdown, NSW 2006, Australia. \protect E-mail: athman.bouguettaya@sydney.edu.au

\IEEEcompsocthanksitem Yun Yang is with the Department of Computing Technologies, Swinburne University of Technology, Melbourne, VIC 3122, Australia. \protect E-mail: yyang@swin.edu.au.}
 \thanks{Manuscript received XXX XXX, 2023; revised XXX XXX, 2023.}   
}

\markboth{IEEE Transactions of Service Computing,~Vol.~xx, No.~x, August~2023}%
{Bo Li \MakeLowercase{\textit{et al.}}: EdgeDis: Enabling Fast, Economical, and Reliable Data Dissemination for Mobile Edge Computing}

\IEEEtitleabstractindextext{%
\begin{abstract}
Mobile edge computing (MEC) enables web data caching in close geographic proximity to end users. Popular data can be cached on edge servers located less than hundreds of meters away from end users. This ensures bounded latency guarantees for various latency-sensitive web applications. However, transmitting a large volume of data out of the cloud onto many geographically-distributed web servers individually can be expensive. In addition, web content dissemination may be interrupted by various intentional and accidental events in the volatile MEC environment, which undermines dissemination efficiency and subsequently incurs extra transmission costs. To tackle the above challenges, we present a novel scheme named EdgeDis that coordinates data dissemination by distributed consensus among those servers. We analyze EdgeDis's validity theoretically and evaluate its performance experimentally. Results demonstrate that compared with baseline and state-of-the-art schemes, EdgeDis: 1) is 5.97x - 7.52x faster; 2) reduces dissemination costs by 48.21\% to 91.87\%; and 3) reduces performance loss caused by dissemination failures by up to 97.30\% in time and 96.35\% in costs.
\end{abstract}

\begin{IEEEkeywords}
mobile edge computing, data caching, data dissemination, distributed consensus, efficiency, and reliability
\end{IEEEkeywords}}

\maketitle

\IEEEdisplaynontitleabstractindextext

%
\IEEEpeerreviewmaketitle

\IEEEraisesectionheading{
\section{Introduction}
\label{Sec:Introduction}}

\IEEEPARstart{M}{obile} edge computing (MEC) has been widely acknowledged as a key enabler technology for latency-sensitive applications in 5G. By deploying edge servers to locations geographically close to end users, e.g., base stations and access points, MEC provides app vendors with a virtual edge caching system comprised of computing and storage resources at the edge network~\cite{9155365}. As the physical distance between content and end users is minimized, i.e., usually within hundreds of meters \cite{zhao2022joint}, service latency can be drastically reduced~\cite{li2021cooperative}. This benefits many modern and futuristic online applications~\cite{gao2020rethinking}, especially latency-sensitive applications, such as AR/VR (augmented/virtual reality) applications and online gaming. App vendors can proactively cache popular data (including applications and the data involved) on distributed edge servers to serve nearby users with minimal service latency~\cite{xia2020online}. There is a series of recent studies on edge data caching~\cite{9155233,glasbergen2020chronocache,xia2021cost}, which provides different caching strategies to help app vendors achieve different objectives. 

To implement a caching strategy in an edge caching system, original data must be disseminated from the app vendor's cloud server to the target edge servers for caching. However, edge servers are highly distributed and the total number of replicas is usually large \cite{he2019game}. Data transmission from the cloud to massive edge servers individually over the Internet may take a long time to complete. In addition, it is often not economic. For example, Amazon Web Services charges \$0.05-\$0.09 for every GB of data transferred out of its S3 storage to the Internet\footnote{https://aws.amazon.com/s3/pricing/}. This edge data dissemination (EDD) problem must be solved properly to enable edge caching in practice.

\begin{figure*}[tbp]
\centering
    \centering
    \includegraphics[width=0.7\linewidth]{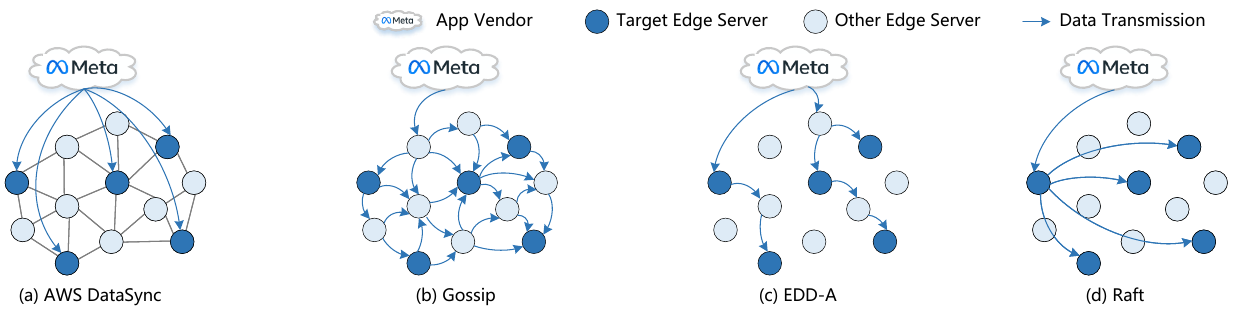}
    \vspace{-1.5em}
    \caption{Exemplar Web Content Dissemination Schemes}    
    \vspace{-1.5em}
    \label{fig:motivation}
\end{figure*}

\smallskip
\textbf{EDD Scenario.} The application scenario of EDD is defined as follows. In a specific geographic area, a set of networked \textit{edge servers}\footnote{We refer to the web servers running on edge servers simply as edge servers hereafter for ease of exposition.} constitute an edge caching system. To implement a data caching strategy, an app vendor needs to disseminate a data item $d$ to a set of $n$ target edge servers in the system. This system differs from conventional cloud caching systems like content delivery networks (CDNs) with three unique characteristics. First, data demands at the network edge are highly dynamic~\cite{li2021Inspecting}. Therefore, EDD must be efficient to respond to data demands promptly. Second, a large number of edge servers (cache nodes) are distributed in the system far away from the cloud~\cite{li2021cooperative}. Transmitting data directly from the cloud to edge servers individually is not economical. Third, edge servers are geographically and highly distributed \cite{zhao2021joint}. They are subject to failures, which may slow down, suspend, or  interrupt the EDD process.

EDD schemes can be designed based on existing data transmission protocols for distributed systems. However, they each suffer from specific limitations.

\smallskip
\textbf{DataSync~\cite{datasync2018AWS}:} DataSync is Amazon's end-to-end data transfer service. Under a DataSync-like EDD scheme, a replica of $d$ needs to be transmitted from the cloud server to each of the $n$ target edge servers individually, as shown in \textbf{Fig. \ref{fig:motivation}(a)} where $n = 5$. This can incur excessive traffic overheads over the backhaul network, which conflicts with MEC's fundamental pursuit of backhaul traffic reduction~\cite{li2020Auditing}. Transmitting a large volume of data out of the cloud also incurs high transmission costs~\cite{xia2021cost}, which the app vendor may not be able to afford.

\smallskip
\textbf{Gossip~\cite{boyd2006randomized}:} The Gossip protocol is widely used in distributed and networked systems to share information across a large number of nodes \cite{boyd2006randomized}. Compared with DataSync, Gossip incurs much less backhaul network traffic. However, as shown in \textbf{Fig. \ref{fig:motivation}(b)}, the peer-to-peer data transmission between edge servers incurs enormous communication overheads between edge servers. 

\smallskip
\textbf{EDD-A~\cite{xia2021cost}:} It formulates EDD strategies that balance the data transmission costs and time consumption. As shown in \textbf{Fig. \ref{fig:motivation}(c)}, it finds a Steiner tree that contains the cloud server as the root node and the $n$ target edge servers as leaf or intermediate nodes. Data item $d$ is iteratively transmitted along the paths from the cloud server to target edge servers. EDD-A usually suffers from single-node failures, which is common in the volatile MEC environment comprised of unreliable edge servers and unstable networks \cite{yuan2021CoopEdge}.

\smallskip
\textbf{Raft~\cite{ongaro2014Raft}:} As a well-known distributed consensus protocol, it helps fallible participants in a system achieve consensus eventually. As shown in \textbf{Fig. \ref{fig:motivation}(d)}, an edge server is elected as the leader to receive $d$ from the cloud and forward it to the other target edge servers. However, it is inefficient only the data dissemination is controlled by the leader, which bottlenecks the overall performance easily. Second, Raft employs sophisticated mechanisms to ensure strong consistency between multiple replicated state machines which are not necessary in EDD scenarios where the order of data transmissions is not mandatory.

\smallskip
\textbf{Contributions.} This paper presents EdgeDis, a novel scheme that enables fast, economical, and reliable edge data dissemination. To the best of our knowledge, EdgeDis is the first EDD scheme systematically designed to facilitate edge data dissemination. Its features and main contributions are summarized as follows.

\begin{itemize}[leftmargin=*]

    \item To improve data dissemination efficiency, EdgeDis enables two-level data transmission concurrency. On the first level, data item $d$ is partitioned into multiple data blocks which are transmitted from the cloud to multiple \textit{entry edge servers} in the edge caching system simultaneously. On the second level, upon the receipt of a data block $d_i$, an entry edge server disseminates replicas of $d_i$ to all the other edge servers in the system in parallel. 
    
    \item To alleviate bandwidth bottlenecks, EdgeDis selects edge servers with high outbound bandwidths as entry edge servers. (\cref{subSec:Reliable_AppVendorInteraction})

    \item To alleviate the impact of slow edge servers on the dissemination efficiency, EdgeDis enables early commitment, i.e., an entry edge server commits the dissemination after the majority of edge servers have received the data. (\cref{subSec:Reliable_AppVendorInteraction})

    \item To economically disseminate data, each data under EdgeDis is only transmitted once from the cloud to the edge caching system, unless transmission failure occurs. This minimizes the traffic overheads incurred over the backhaul network and thus reduces the cost ~\cite{xia2021cost}. 
    
    \item To ensure reliable dissemination and tackle the challenges caused by communication failures and/or edge server failures, EdgeDis elects an \textit{coordinator} to check the data dissemination status and supply target edge servers with missing data blocks. This ensures that every target edge server will receive an entire replica of $d$ eventually. Notably, the supplement of missing data blocks is performed within the system, which minimizes the extra transmission costs incurred. 
    
    \item We propose a novel coordinator election and maintenance mechanism to ensure coordinator uniqueness. Besides, it also ensures system stability upon coordinator failures. 
    
    \item We analyze the validity of EdgeDis theoretically and implement a prototype of EdgeDis to experimentally demonstrate its performance over baseline and state-of-the-art EDD schemes in a testbed comprised of 128 edge servers. 
\end{itemize}

This paper is structured as follows. Section \ref{Sec:Background_Motivation} reviews the backgrounds and motivates this research.  Section \ref{Sec:Reliable_EDD} depicts EdgeDis's data dissemination processes in detail. Section \ref{Sec:CoordinatorElection} introduces EdgeDis's coordinator election process. Section \ref{sec:Validity_analysis} analyzes EdgeDis's validity and Section \ref{Sec:Experiments} experimentally evaluates EdgeDis's performance.  Finally, Section \ref{Sec:Conclusion} presents the conclusion and discusses future work.


%
\section{Related Work and Motivation}
\label{Sec:Background_Motivation}

\subsection{Data Caching}
Data caching in MEC environment is the basis for applications relying on low-latency data access ~\cite{huang2019fair, xia2020online}, e.g., AR/VR gaming. It has attracted excessive attention from industry and academia. To name a few, Xia et al. proposed an online algorithm to help app vendors find suitable strategies to cache data on distributed edge servers \cite{xia2020online}. Their approach considers data deployment cost, potential migration cost, and impact of quality-of-service (QoS), and minimizes the overall cost. Later, they proposed a similar approach aiming at maximizing the data caching revenue for app vendors \cite{xia2021ol}. Xu et al. took into account the service requirement prediction and proposed E-Cache to find suitable data caching strategies for smart cities, especially the smart vehicles \cite{xu2021edge}.

Many studies have been conducted to find suitable caching strategies for content delivery networks (CDNs). To name a few, Karamchandani et al. studied the hierarchy of CDN with 2 layers of caches. They proposed a new scheme based on the coded caching/computing technique to maximize the communication rate \cite{karamchandani2016hierarchical}.
Zhang et al. designed a double-layer coded caching system for ultra-dense networks (UDNs) \cite{zhang2019double}. They proposed a deep reinforcement learning-based approach to reduce the average delay and power consumption of wireless networks. Wan et al. proposed an approach to minimize the worst-case communication overheads when caching location-based content \cite{wan2022optimal}. However, the problems studied in the above research are different from the problem studies in this paper. Specifically, coded caching encrypts different files to minimize the overall communication overheads when users access different files \cite{cheng2021novel}. In contrast, EdgeDis quickly, economically, and reliably disseminates replicas of one data to multiple edge servers.

\vspace{-1em}
\subsection{Data Dissemination}

Many conventional data transmission approaches can be employed to facilitate data dissemination in the MEC environment. For example, as a well-known peer-to-peer (P2P) data-sharing mechanism, Gossip~\cite{boyd2006randomized} can be used to transfer data between edge servers. However, this will inevitably incur heavy communication overheads over the fronthaul edge server network. The device-to-device (D2D) communication approaches like \cite{feng2014device} used in mobile networks are also useful. However, as edge servers are geographically distributed, merely enabling nearby edge servers to communicate and share data is not useful enough. Similarly, the machine-to-machine (M2M) technique used in sensor networks to exchange data between smart devices is not useful enough due to the performance limitations \cite{verma2016machine}.

To realize a data caching strategy in the MEC environment, app vendors need to transmit their data to target edge servers. This raises the data dissemination problem. Xia et al. investigated this problem and proposed EDD-A to tackle this problem \cite{xia2021cost}.  Later, they extended the data dissemination to dynamic scenarios where a cached data replica can be further moved dynamically. They proposed an online approach based on Lyapunov optimization to solve the problem \cite{xia2022formulating}. However, both studies formulated the data dissemination problem as an optimization problem. They only proposed approaches to find dissemination strategies aiming at reducing data transmission costs. Neither of them has considered the transmission reliability and/or transmission efficiency.

\setlength{\textfloatsep}{4pt}
\begin{table}[bt]
\renewcommand{\arraystretch}{1.1}
    \caption{Limitations of Potential EDD Schemes}   
    \vspace{-1em} 
    \label{Tab:Characteristics}
    \centering
    \scalebox{1}{\begin{tabular}{l|c|c|c}
    \hline
\hline
    Schemes & G1 Efficiency & G2 Economy & G3 Reliability\\
    \hline
        DataSync\cite{datasync2018AWS} & $\times$  & $\times$ & $\times$ \\
        Gossip~\cite{boyd2006randomized} & $\times$  & $\times$ & $\checkmark$ \\
        EDD-A~\cite{xia2021cost} &  $\times$  & $\checkmark$  & $\times$ \\
        Raft~\cite{ongaro2014Raft} & $\times$  & $\checkmark$ & $\checkmark$\\
    \hline
\hline
    \end{tabular}}
\end{table}

\subsection{Motivation}
Given an edge data caching strategy, data can be disseminated from the cloud to target edge servers following different protocols, e.g., DataSync~\cite{datasync2018AWS}, Gossip~\cite{boyd2006randomized}, EDD-A~\cite{xia2021cost}, Raft~\cite{ongaro2014Raft}. However, these protocols are either too slow or too expensive in EDD scenarios (\cref{Sec:Introduction}). A practical EDD scheme must achieve the following three performance goals:
\begin{itemize}
    \item \textbf{G1 Efficiency:} Edge data dissemination must be completed rapidly to enable low service latency and cope with dynamic data demands at the network edge.
    \item \textbf{G2 Economy:} Edge data dissemination must not incur excessive traffic overheads over the backhaul network or the edge network. Based on the pay-as-you-go pricing model~\cite{wang2021eihdp}, traffic overhead reduction immediately lowers monetary costs for app vendors.
    \item \textbf{G3 Reliability:} In the volatile MEC environment, various events can lead to data dissemination failures. A practical EDD scheme must adapt to data dissemination failures during the EDD process.
\end{itemize}

None of the existing data transmission schemes (\cref{Sec:Introduction}) can achieve all the above performance goals. Their limitations are summarized in \textbf{Table \ref{Tab:Characteristics}}. Symbol $\checkmark$ means the objective achieved and $\times$ means the objective not achieved.

From the system perspective, when a data item $d$ is successfully disseminated to all the target edge servers, these edge servers are consistent in terms of the data. This aligns with the concept of distributed consensus. In fact, many consensus-based schemes have been designed and employed in industry to ensure data consistency in distributed database/file systems~\cite{ skrzypczak2020rmwpaxos, tollman2021epaxos}, such as Spanner~\cite{corbett2013spanner}, ZooKeeper~\cite{hunt2010zookeeper}, and Chubby~\cite{burrows2006chubby}. They have applied Paxos~\cite{lamport2001paxos} or Raft~\cite{ongaro2014Raft} to facilitate distributed consensus. A main advantage is the ability to tolerate faults. It can be leveraged by EDD schemes to adapt to data dissemination failures\footnote{In this study, we consider temporary dissemination failures, not permanent failures that impact the system continuously.}. Parallelization, which has been widely used to improve the efficiency of computer systems~\cite{MLSYS2022_d96409bf,9472938}, is a potential solution to alleviating the performance bottleneck created by single leader under leader-based distributed consensus protocols like Raft \cite{ongaro2014Raft}.

\begin{figure*}
  \centering
        \includegraphics[width=0.7\linewidth]{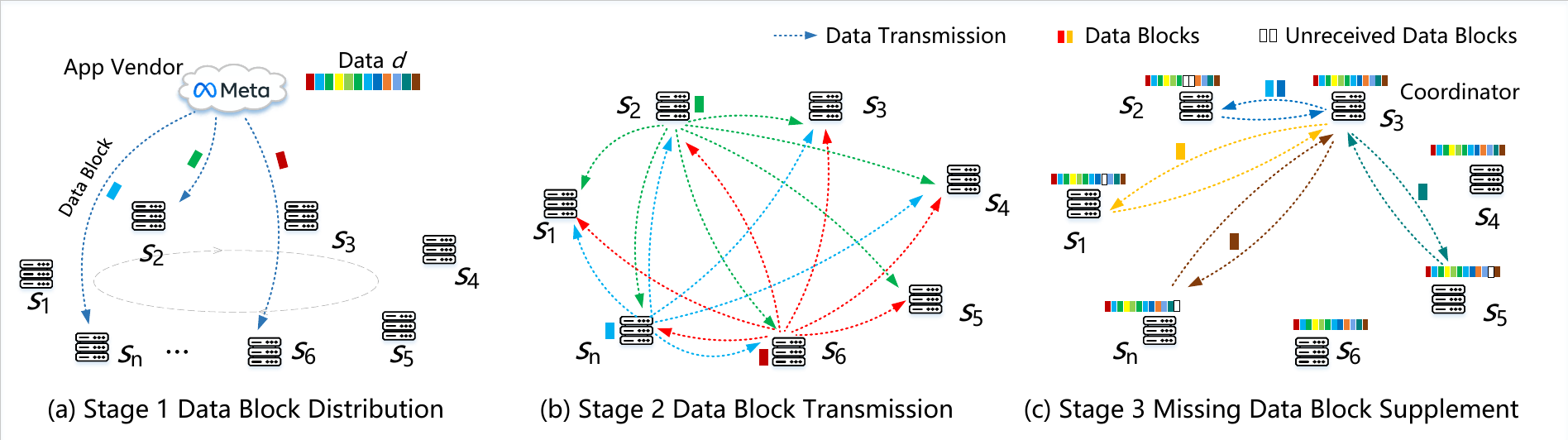}
    \vspace{-1em}
        \caption{Overview of EdgeDis's Data Dissemination Process. }        
    \vspace{-1em}
        \label{Fig:EdgeDis_Scheme}
\end{figure*}

The explosive increase in Internet traffic puts an increasing burden on the backhaul network \cite{xia2021ol}. Routing data over the edge network instead of the backhaul network is a promising solution \cite{landa2021staying, yap2017taking, schlinker2017engineering}. Thus, the key to achieving G2 Economy is to minimize the volume of data transmitted out of the cloud during the EDD process. Raft is not likely to struggle with achieving G2 Economy, as data $d$ is transmitted only once to the leader who forwards $d$ to other target servers over local edge network.

When dissemination failures occur during the EDD process, some target edge servers may not receive all the data blocks. A simple solution is for these edge servers to fetch the missing data blocks from the cloud. This will inevitably incur extra traffic out of the cloud, which undermines the effort to achieve G2 Economy. In fact, if internal dissemination failures can be handled within the system, there is no need to fetch missing data blocks from the cloud, and thus G1 Efficiency can be achieved. The unique challenges in EDD motivate us to design EdgeDis with the aim of achieving all three performance goals. Under EdgeDis, $d$ is partitioned into multiple data blocks. Each of these data blocks will be sent to an entry edge server for transmission to target edge servers\footnote{EdgeDis makes no assumption on specific transport protocols for the transmission of these data blocks.}. In this way, parallelism is applied to the dissemination of $d$ into and within the edge caching system. This achieves G1 Efficiency. Each data block is transmitted from the cloud to the system only once unless transmission failures occur. This minimizes the backhaul network traffic and achieves G2. EdgeDis elects an edge server within the system to supply missing data blocks. This confines the traffic to the system and allows EdgeDis to achieve G3 Reliability without undermining G2.


%
\section{Decentralized Edge Data Dissemination}
\label{Sec:Reliable_EDD}

Now we introduce EdgeDis, considering the dissemination of a data item $d$ to $n$ edge servers. The dissemination of multiple data items can be carried out individually and concurrently under EdgeDis.

EdgeDis goes through three asynchronous stages to disseminate data $d$, as shown in \textbf{Fig. \ref{Fig:EdgeDis_Scheme}}. Note that $d$ is transmitted by blocks under EdgeDis, i.e., $\{d_1, d_2, ...\}$. In Stage 1, the app vendor transmits each data block of $d$ to a specific \textit{entry edge server}. Different data blocks are transmitted in parallel to improve EdgeDis's data dissemination efficiency. In Stage 2, each entry edge server distributes the data block received from app vendor to all the other edge servers in the system. Considering that the transmission can be interrupted, especially when a large number of data blocks are transmitted concurrently, a coordinator is elected in Stage 3 from edge servers. It infers the dissemination status by monitoring data block forward progress in the system. Then, it supplements missing data blocks caused by dissemination failures. This avoids retrieving data from remote cloud. Note that these stages are performed asynchronously. For example, while an entry edge server receives a new data block from app vendor's cloud in Stage 1, it can continue to transmit data blocks received before to the target edge servers in Stage 2. In the meantime, the coordinator can supply missing data blocks in Stage 3. This novel design allows EdgeDis to achieve superior performance, which will be evaluated in Section \ref{subSec:Experimets_ExperimentResults}.

During the data dissemination process, an edge server under EdgeDis may serve two roles at the same time: \textit{sender} and \textit{receiver}. 
\begin{itemize}
    \item \textbf{Sender} - A sender is an entry edge server in the system that receives data blocks from the cloud and transmits them to the other edge servers.
    \item \textbf{Receiver} - A receiver receives data blocks from senders and responds to senders to confirm the receipts of their data blocks.
\end{itemize}

\subsection{Stage 1: Data Block Distribution from Cloud to Entry Edge Servers}
\label{subSec:Reliable_AppVendorInteraction}

Following the common design of many consensus-based systems~\cite{zhou2019highly, sonbol2020edgekv}, EdgeDis assigns a unique and consecutive integer ID to each data block in data $d$ before dissemination. Then, the app vendor's cloud server $\mathbi{cs}$ disseminates the data blocks of $d$ to entry edge servers in the system. Take \textbf{Fig. \ref{Fig:EdgeDis_Scheme}(a)} as an example. Cloud server $\mathbi{cs}$ sends data block $d_1$ to edge server $s_6$, $d_2$ to $s_n$, etc. After sending a data block $d_i$ to edge server $s_j$, $s_j$ becomes a \textit{sender} and transmits $d_i$ to the other receivers. Under EdgeDis, $\mathbi{cs}$ can send multiple data blocks continuously to different entry edge servers simultaneously without waiting for their confirmed receipts of previous data blocks. This is a key to EdgeDis's pursuit of G1 Efficiency. 

Considering that edge servers are highly heterogeneous in network capacities and workloads~\cite{yao2017heterogeneous, pasteris2019service, gai2017sa}, entry edge servers with limited capacities, e.g., constrained outbound bandwidth, can throttle the EDD process significantly. Fortunately, according to the \textit{de facto} MEC standards~\cite{kekki2018mec, sabella2019developing}, app vendors can retrieve specific information about the MEC environment, e.g., edge servers' capacities and base stations' channel conditions. This allows EdgeDis to choose edge servers with high network capacities as entry edge servers. Entry edge servers transmit received data blocks to target edge servers, which is bandwidth-intensive. Thus, EdgeDis chooses edge servers with the highest outbound bandwidth as entry edge servers to alleviate performance bottlenecks. 

There is a trade-off when determining the number of entry edge servers. For example, more entry edge servers can minimize the workload of the entry edge servers and improve dissemination efficiency. However, this requires more bandwidth capacities on receivers and the cloud server, as those entry edge servers transmit data in parallel. In contrast, employing fewer entry edge servers can reduce the burden on the cloud server and receiver edge servers, but undermines the dissemination efficiency. In practice, the number of entry edge servers can be empirically determined to make full use of available network capacity of app vendor's cloud server and those edge servers in the system. In the experiments, we choose the top 25\% edge servers in the system with the highest available bandwidth as entry edge servers. This reduces the average workload of entry edge servers, improves data dissemination efficiency, and does not exhaust incoming bandwidth on receiver edge servers.

The cloud server $\mathbi{cs}$ employs an array $status$ to record the data block distribution progress. For example, $\mathbi{cs}.status[i]=1$ indicates that data block $d_i$ has been successfully transmitted to the corresponding entry edge server. The array could be further extended to a 2D array to facilitate the dissemination of multiple data. The data dissemination completes when the receipts of all the data blocks are confirmed by the senders with \textit{data block distribution completion messages}. If $\mathbi{cs}$ does not receive a confirmation message from $s_j$ before \textit{distribution timeout}, it resends $d_i$ to another available edge server for distribution. In this way, a data block needs to be transmitted only once from $\mathbi{cs}$, except when failures occur. This significantly reduces the backhaul network traffic and allows EdgeDis to achieve G2.

\smallskip
\textbf{Step 1. Data Block Distribution Message Generation.} To transmit a data block $d_i$ to edge server $s_j$, cloud server $\mathbi{cs}$ sends a message $mes_{cs}$ to $s_j$ with data block $d_i$ with data block ID, i.e., $mes_{cs}.dataBlock=d_i$ and $mes_{cs}.dataBlockId=i$. Then, $s_j$ responds with a data block distribution completion message to confirm the receipt of $d_i$. If $\mathbi{cs}$ does not receive a response from $s_j$ by the distribution timeout, it randomly sends $d_i$ to another server.

\smallskip
\textbf{Step 2. Data Block Distribution Completion Message Generation.} When edge server $s_j$ receives a data block distribution message $mes_{cs}$, it saves the corresponding data block $d_i$ in $mes_{cs}$ to its local storage. Then, it transmits replicas of $d_i$ to all the other edge servers (\cref{subSec:DataExchange}). When the \textit{majority}\footnote{Same to Raft~\cite{ongaro2014Raft}, in this research, given $n$ target edge servers, a majority of edge servers consists of at least $\lceil (n+1)/2\rceil$ edge servers.} of edge servers have acknowledged the receipt of $d_i$, $s_j$ sends a \textit{data block distribution completion message} to $\mathbi{cs}$ to confirm the completion of the dissemination of $d_i$. 

\subsection{Stage 2: Data Block Transmission between Edge Servers}
\label{subSec:DataExchange}

As shown in \textbf{Fig. \ref{Fig:EdgeDis_Scheme}(b)}, when an entry edge server $s_j$ receives a data block $d_i$ from $\mathbi{cs}$, it becomes a sender and transmits $d_i$ via \textit{data block transmission messages} to the other edge servers in the system. Each of these other edge servers, denoted by $s_k$, responds to $s_j$ with a \textit{data block receipt message} to confirm the receipt of $d_i$. Due to the inherent characteristics of distributed systems, data transmission between edge servers is volatile and may fail~\cite{ rafique2020complementing}. Edge server $s_j$ will resend $d_i$ to $s_k$ if it does not receive a response from $s_k$ before the \textit{transmission timeout}. In this way, $s_k$ can receive $d_i$ after it recovers from a temporary failure. When the majority of edge servers in the system have received $d_i$, entry edge server $s_j$ responds to $\mathbi{cs}$ with a data block distribution completion message to confirm the completion of the dissemination of $d_i$. In the meantime, $s_j$ will continuously transmit $d_i$ to the remaining edge servers. This avoids delays caused by dissemination failures or slow receivers in the system. The workflows followed by the senders and receivers are shown in \textbf{Appendix A Pseudocodes 1} and \textbf{2}, respectively. This stage consists of three steps.

\smallskip
\textbf{Step 1. Data Block Transmission Message Generation.} The sender $s_j$ generates a data block transmission message $mes_s$ and sends it to all the other edge servers in the system. This message contains data block $d_i$ and its data block ID, i.e., $mes_s.dataBlock = d_i$ and $mes_s.dataBlockId = i$. 

\smallskip
\textbf{Step 2. Data Block Receipt Message Generation.} Upon the receipt of a data block transmission message $mes_s$ from $s_j$, receiver $s_k$ stores data block $d_i$ in its local storage and updates its data caching status by setting $s_k.dataBlockStatus[i]=1$. If $d_i$ has the largest ID over all the data blocks cached on $s_k$, $s_k$ will update $s_k.maxBlockId=i$. Then, it generates a \textit{data block receipt message} $mes_r$ to confirm the receipt of $d_i$ with sender $s_j$.

\smallskip
\textbf{Step 3. Data Block Transmission Progress Inspection.} Upon the receipt of a data block receipt message $rec_r$ from receiver $s_k$, sender $s_j$ records the status of the transmission of $d_i$ to $s_k$. When data block $d_i$ is successfully transmitted to the majority of the edge servers in the system, it notifies the cloud server $\mathbi{cs}$ of the completion of $d_i$. This also indicates that $s_j$ is ready to receive the next data block from $\mathbi{cs}$. In the meantime, $s_j$ continues to transmit $d_i$ to the remaining receivers that have not received $d_i$. If $s_j$ does not receive a response from a receiver $s_k$ by the transmission timeout, it resends $mes_s$ at its next attempt. This allows EdgeDis to adapt to internal dissemination failures caused by slow nodes and network congestion efficiently and economically, compared with resending data from $cs$ upon dissemination failures.

\subsection{Stage 3: Missing Data Block Supplement}
\label{subSec:MissingDataRecovery}

While dissemination failures between senders and receivers are handled in Stage 2, an entry edge server may fail before successfully transmitting a data block $d_i$ to all the receivers. To adapt to such sender failures, EdgeDis elects an edge server as \textit{coordinator} to supply missing data blocks to edge servers.

To be able to supply any missing data blocks, the coordinator collects all the data blocks distributed to the system through periodical heartbeat requests and responses. This also helps receivers determine whether any data blocks are missing. Note that an individual edge server does not know how many non-duplicate data blocks will eventually be received in total for building data $d$. Thus, the coordinator is indispensable. Knowing that a data block is missing, a receiver can retrieve that data block from the coordinator, as shown in \textbf{Fig. \ref{Fig:EdgeDis_Scheme}(c)}. The coordinator holds its position for a time period, referred to as a \textit{term}. Terms are numbered with consecutive integer IDs, similar to Raft~\cite{ongaro2014Raft}. When the current coordinator fails, the other edge servers in the system may start their individual election processes to contend for the coordinator role (\cref{Sec:CoordinatorElection}). The new coordinator will resume the supplement of missing data blocks. In this way, the edge servers constitute an autonomous system immune to coordinator failures.

\smallskip
\textbf{Step 1. Data Block Collection.}
Under EdgeDis, the coordinator $\mathbi{c}$ maintains its role by periodically sending heartbeats to and receiving heartbeat receipt messages from all the followers, the underlying mechanism is similar to Raft~\cite{ongaro2014Raft}. However, different from Raft, a coordinator under EdgeDis reuses those heartbeat messages to inspect the data block possession status of all followers. Then, $\mathbi{c}$ collects missing data blocks according to such status. The workflows are depicted by \textbf{Pseudocodes 3} and \textbf{4 in Appendix A}.

First, when sending out a \textit{heartbeat message}, the coordinator $\mathbi{c}$ attaches a variable $\mathbi{c}.maxBlockId$ to the message. The variable $maxBlockId$ is the maximum ID of all the data blocks possessed by $\mathbi{c}$. The heartbeat message, denoted as $mes_h$, is sent to all the other edge servers in the system.

Second, upon the receipt of $mes_h$, a follower $\mathbi{f}$ checks its data blocks according to the received $mes_h.maxBlockId$.  Note that $\mathbi{f}$ has a vector $\mathbi{f}.dataBlockStatus[]$ indicating which data blocks have been received successfully and a variable $maxBlockId$ recording the maximum ID of received data blocks. Hence, if $mes_h.maxBlockId > \mathbi{f}.maxBlockId$, all the data blocks with IDs from $\mathbi{f}.maxBlockId+1$ to $mes_h.maxBlockId$ are missing. Besides, if the value of $\mathbi{f}.dataBlockStatus[i] (i<\mathbi{f}.maxBlockId)$ is 0, the $i$th data block is missing.  Please note that a data block may be delayed by temporary network congestion. In practice, to solve this issue, the coordinator $\mathbi{c}$ usually waits for a short time before it begins to supplement missing data blocks to follower $\mathbi{f}$, e.g., $2\Delta t$ where $\Delta t$ indicates the network diameter. In this case, if a congested data block is eventually received within this time, it can be updated by $\mathbi{f}$'s heartbeat receipt message. Now, follower $\mathbi{f}$ updates its $\mathbi{f}.maxBlockId = mes_h.maxBlockId$ to record the maximum data block ID perceived currently. After that, $\mathbi{f}$ sends a \textit{heartbeat receipt message} to $\mathbi{c}$. A heartbeat receipt message includes $\mathbi{f}.maxBlockId$ to notify $\mathbi{c}$ of the maximum data block ID it perceived. It also includes an array $\mathbi{f}.missBlockId[]$ indicating which data blocks are missing from $\mathbi{f}$'s local storage.

Third, by inspecting all heartbeat receipt messages, coordinator $\mathbi{c}$ can find out how many data blocks have been distributed to the system, which is indicated by the maximum value of all received $maxBlockId$ variables. It can also find out which edge servers possess the data blocks missing from its local storage. Then, as shown in Pseudocode 3 in Appendix A, $\mathbi{c}$ sends a \textit{data block request} to fetch its missing data blocks from these edge servers. Upon the receipt of a data block request, the receiver returns a \textit{data block response} to $\mathbi{c}$ with the required data blocks. This helps $\mathbi{c}$ collect all data blocks distributed in the system.

\smallskip
\textbf{Step 2. Data Block Supplement.}
After receiving a new heartbeat receipt message $mes_h^f$ from follower $\mathbi{f}$, coordinator $\mathbi{c}$ can easily determine which data blocks are missing on $\mathbi{f}$ based on the received $mes_h^f.missBlockId[]$.  This allows coordinator $\mathbi{c}$ to proactively transmit the missing data blocks to $\mathbi{f}$ by sending a data block supplement message to $\mathbi{f}$ with missing data blocks attached. Then, upon the receipt of these data blocks, $\mathbi{f}$ stores them into its local storage. It is unnecessary to confirm the receipt of the data block supplement message. As introduced in Step 1, the coordinator $\mathbi{c}$ periodically sends heartbeats to $\mathbi{f}$, which helps collect $\mathbi{f}$'s data block possession status. If the vector $missBlockId[]$ in $\mathbi{f}$'s next heartbeat receipt message is empty, i.e., $mes_h^f.missBlockID=null$, the missing data blocks have been successfully supplemented. Otherwise,  $\mathbi{c}$ supplements again the data blocks according to the updated $missBlockID[]$.

\textit{\textbf{Remark 1: } Under Raft, EDD relies on an individual leader. A leader failure will suspend the EDD process immediately until a new leader is elected. Unlike Raft, EdgeDis selects multiple entry edge servers to transmit data blocks. The coordinator is elected to supply receivers with missing data blocks. It will not supplement a data block unless the corresponding entry edge server fails. Thus, the coordinator's workload is relatively light. In addition, if the coordinator is exhausted and cannot respond in time, EdgeDis will dynamically elect a new coordinator to take over the responsibility. Please note that coordinator failures may delay the supplement of missing data blocks, but will not suspend the entire EDD process, which is better than Raft.}

\textit{\textbf{Remark 2}:  The coordinator communicates with every edge server in the system by heartbeats to periodically check the dissemination status. Given $n$ edge servers in the system, there are $n$ heartbeats per time slot. Thus, the overall communication overheads are determined by $n$, the time interval $ti$ for sending heartbeats, and the overall time consumed for data dissemination $T$.  A heartbeat message is 12 Bytes. A heartbeat receipt message is at least 12 Bytes and can be longer, depending on the number of missed data blocks on each edge server. Thus, the overheads are at least $24\times n\times T/ti$ Bytes. Note that as the coordinator periodically supplements missing data blocks, the number of missed ones could not be great. Please see Section \ref{subsubsec:systemOverhead} for the experimental evaluation.}


%
\section{Coordinator Election}
\label{Sec:CoordinatorElection}
A coordinator is employed to monitor data dissemination status on edge servers and supplement missing data blocks if any. Inspired by distributed consensus protocols like Paxos \cite{lamport2001paxos} and Raft \cite{ongaro2014Raft}, EdgeDis allows only one coordinator in the system. This simplifies the system design and reduces relevant network traffic while fulfilling the functionality requirements. A new coordinator will be elected when the system starts or the current coordinator fails. To facilitate coordinator election, edge servers may be in one of the three states.  
\begin{itemize}
    \item \textbf{Coordinator} - Only one coordinator in the system at any time. It is responsible for supplying edge servers with missing data blocks.
    \item  \textbf{Follower} - When there is a coordinator in the system, all the other edge servers are in the follower state. When a sender fails, a follower edge server can retrieve the missing data blocks from the coordinator.
    \item  \textbf{Candidate} -  An intermediate state between coordinator and follower. When a follower starts its election process, it becomes a candidate.
\end{itemize}

When data dissemination begins, all the edge servers are followers. Each follower $\mathbi{f}$ starts a timer with a \textit{coordinator timeout} randomly set between $t$ and $2\times t$. When $\mathbi{f}$ does not receive any heartbeats from the coordinator when its \textit{coordinator timeout} elapses, it becomes a candidate and starts its coordinator election process to contend for the coordinator role. Otherwise, it resets the coordinator timeout. A candidate becomes the coordinator if it wins the election by obtaining support from the majority of edge servers. When a candidate receives a heartbeat from the coordinator during the election process, it becomes a follower immediately. The coordinator becomes a follower when it fails or perceives the existence of a newer coordinator with a larger term ID. In general, the election process consists of three main steps.

\smallskip
\textbf{Step 1. Vote Request Generation.} Candidate $\mathbi{c}_f$ increases its term ID by 1, then generates and sends a \textit{vote request} $req$ to all the other edge servers, as shown in \textbf{Pseudocode 5 in Appendix A}. The request $req$ contains the number of data blocks $\mathbi{c}_f$ has and its term ID.

\smallskip
\textbf{Step 2. Vote Response Generation.} Each edge server $\mathbi{f}$ responds to $\mathbi{c}_f$'s vote request $req$, as shown in \textbf{Pseudocode 6 in Appendix A}. Edge server $\mathbi{f}$ supports $\mathbi{c}_f$ if three conditions are fulfilled: 1) $\mathbi{f}$ does not have more data blocks than $\mathbi{c}_f$; and 2) $\mathbi{f}$'s term ID is not larger than $\mathbi{c}_f$'s, i.e., $req.currentTermId \geq \mathbi{f}.termId$; and 3) it has not supported any other candidate whose Term ID is equal to $req.currentTermId$. For example, if $\mathbi{f}$ has supported a candidate $\mathbi{c}_f$ whose term ID is 5, it will reject all the other candidates' vote requests whose term IDs are also 5. Please note that $\mathbi{f}$ can continue to support a candidate whose term ID is larger than 5. In other words, given a specific term ID value, $\mathbi{f}$ can support at most one candidate with this Term ID. 

Condition 1) ensures that a new coordinator has more data blocks than the majority of edge servers. This can minimize the time for the new coordinator to complete the data block collection (see Section \ref{subSec:MissingDataRecovery}). Condition 2) ensures that a candidate with a small term ID cannot contend for the coordinator role. Inequality $\mathbi{f}.termId>req.currentTermId$ indicates that either there is a new coordinator in the system or another edge server with a larger term ID that has started its election process earlier. In such cases, $\mathbi{c}_f$ cannot compete for the coordinator role. Condition 3) ensures that only one coordinator will be elected with any term ID. 

If all the three constraints are fulfilled, $\mathbi{f}$ updates its term ID with $\mathbi{c}_f$'s term ID found in $req$, and offers its support to $\mathbi{c}_f$ by setting $res.supported=true$, as shown in Pseudocode 6. Otherwise, it sets $res.supported=false$. Then, it sends $res$ to $\mathbi{c}_f$. Note that, $\mathbi{f}$'s term ID is attached to $res$ by setting $res.currentTermId = \mathbi{f}.termId$ so that $\mathbi{c}_f$ can find out at Step 3 whether $\mathbi{f}$ has a bigger term ID.

\smallskip
\textbf{Step 3. Vote Counting.} After receiving a set of \textit{vote responses}, as shown in Pseudocode 5, candidate $\mathbi{c}_f$ inspects if the majority of the edge servers (including itself) in the system support it in the election. If so, it becomes the new coordinator immediately and begins to send heartbeats periodically to all the other edge servers to declare its coordinator role. Otherwise, it repeats the election process after timeout, until obtains support from the majority or a new coordinator is elected. If $\mathbi{c}_f$ finds a bigger term ID, it terminates its election process and becomes a follower immediately.


\section{Validity Analysis}
\label{sec:Validity_analysis}
Now we theoretically analyze EdgeDis's validity, including the convergence of coordinator election, coordinator uniqueness, and system reliability upon edge server failures.

\subsection{Coordinator Election Convergence}
\label{subSection:Convergence_CoordinatorElection}

A candidate edge server $\mathbi{c}_f$ cannot be a coordinator unless it obtains majority support in the system during the election process. However, when multiple candidates compete at the same time, vote splitting may occur and no candidate can obtain majority support. EdgeDis addresses this issue via randomization, similar to Raft~\cite{ongaro2014Raft}. Specifically, instead of commencing the election process immediately at the coordinator timeout, an edge server randomly waits for a time period between $t$ and $2t$ before starting its election process. This mechanism reduces the possibility that multiple edge servers contend for the coordinator role at the same time. If vote splitting still happens, the candidate will wait for another random time period between $t$ and $2t$ before its next attempt. The election process will converge and one coordinator will be elected eventually. The convergence will be experimentally studied in Section \ref{subsubsec:parameterImpacts}.

\subsection{Coordinator Uniqueness}

\label{subSection:Safety_CoordinatorElection}
Coordinator uniqueness demands that there is at most one active coordinator in the system at any time. Now we analyze that EdgeDis fulfills this demand.  Let us assume that $s_i$ and $s_j$ have commenced their own election processes and become candidates. Note that an edge server increases its Term ID by 1 when timeouts and becomes a candidate.  There are five potential cases. 

\smallskip
\textbf{Case 1: Two candidates $s_i$ and $s_j$ have the same term ID and only one is elected.} According to the settings introduced in Section \ref{Sec:CoordinatorElection}, each edge server can support only one candidate for each term ID, then at most one candidate, i.e., either $s_i$ or $s_j$ can receive support from majority follower edge servers. In this case, only one candidate who sends out vote requests faster will become the new coordinator.

\smallskip
\textbf{Case 2: Two candidates $s_i$ and $s_j$ have the same term ID and the votes are even.} This case is that $s_i$ and $s_j$ receive equal numbers of votes from other edge servers. This phenomenon is called \textit{vote split}. The vote split also means neither of them can receive support from the majority of edge servers. Therefore, none of them can be the new coordinator. However, as each candidate edge server has a random timeout threshold between $t$ and $2t$. They will repeat the election processes to contend for the leadership again until one coordinator is successfully elected.

\smallskip
\textbf{Case 3: Multiple candidates with the same term ID competing for leadership.}  The same as the phenomenon discussed in Case 2, when multiple candidates with the same terms ID compete for the leadership, it is probable that there is a vote split. Then, they will repeat the election process again after the timeout. However, as different candidates have different timeout thresholds, they will start the election at different times, which limits the probability of a vote split in the next round \cite{ongaro2014Raft}.

\smallskip
\textbf{Case 4: Two candidates $s_i$ and $s_j$ have different term IDs and only one is elected. } Suppose $s_i$ has a smaller term ID than $s_j$, when a follower edge server receives the vote requests from both $s_i$ and $s_j$ at the same time, it only supports $s_j$. Therefore, only $s_j$ has the opportunity to be supported by the majority of edge servers. Then, $s_j$ will be the new coordinator. 

\smallskip
\textbf{Case 5: Two candidates $s_i$ and $s_j$ have different term IDs and both are elected.} An edge server $s_k$ can support different candidates with different term IDs when receiving vote requests at different times. Suppose that $s_j$'s term ID is larger than $s_i$'s. A potential case is that  $s_i$ just becomes the new coordinator but has not sent its heartbeats to $s_j$. At the same time, as $s_j$ has a greater term ID, it can also receive support from the majority of edge servers and become a coordinator. Therefore, there are two coordinators in the system. Note that as both $s_i$ and $s_j$ are supported by the majority of edge servers, at least one edge server has supported both $s_i$ and $s_j$, say $s_k$. Then, $s_k$'s term ID is equal to $s_j$'s (see Section \ref{Sec:CoordinatorElection}). Candidate $s_i$ will receive this term ID from $s_k$'s heartbeat receipt messages (see Appendix A Pseudocode 4). As shown in Pseudocode 3, $s_i$ becomes a follower immediately once it knows the existence of $s_j$, as its Term ID is less than $s_j$'s. $s_j$ remains as the coordinator in the system. In this case, although there might be more than one candidate in the system, only one can survive eventually.

\subsection{Reliability upon Failures}

\label{subSection:Safety_CoordinatorFailure}

\smallskip
\textbf{Upon Receiver Failures.} A sender $\mathbi{s}$ will keep re-sending a data block to a receiver $\mathbi{r}$ if $\mathbi{r}$ fails temporarily. After $\mathbi{r}$ recovers, it will receive the data block. Thus, a receiver failure will not impact system reliability and the data dissemination will complete eventually.

\smallskip
\textbf{Upon Sender Failure.} If a sender $\mathbi{s}$ fails after transmitting a data block $d_i$ to the majority of edge servers in the system, the coordinator can retrieve $d_i$ from one of these edge servers and continue to transmit $d_i$ to the other edge servers. If $\mathbi{s}$ fails before transmitting $d_i$ to the majority of edge servers, it will not confirm the completion of the transmission with the cloud server. The cloud server will resend $d_i$ to another edge server to be transmitted at the transmission timeout. Therefore, EdgeDis is reliable upon sender failures.

\smallskip
\textbf{Upon Coordinator Failure.} Different from existing distributed consensus like Paxos \cite{lamport2001paxos} and Raft \cite{ongaro2014Raft}  which rely on the coordinator (leader) to perform the data dissemination sequentially, EdgeDis transmits data blocks in parallel by multiple entry edge servers. The coordinator is only responsible for supplying missing data blocks caused by sender failures. A new coordinator will be elected when the current one fails. The recovery of the coordinator will not affect EdgeDis's performance when all entry edge servers work properly. Even if an entry edge fails, as all data blocks are disseminated in parallel, the recovery of the coordinator will slightly affect the data block supplement but will scarcely affect the entire process. Thus, EdgeDis can tolerate coordinator failures.

%
\section{Experimental Evaluation}
\label{Sec:Experiments}
%

\subsection{Experiment Configuration}
\label{sec:experiment_configuration}

\subsubsection{Parameter Settings}
\label{subsubSec:Experimental_Settings}

The coordinator timeout parameter $t$ used for election (\cref{Sec:CoordinatorElection}) is set to 250ms. Every 50ms, the coordinator broadcasts heartbeats in the system to maintain its role. The distribution timeout used for controlling the data block distribution between the cloud and entry edge servers (\cref{subSec:Reliable_AppVendorInteraction}) is set to 300ms. The transmission timeout used for controlling the data block dissemination between sender and receivers (\cref{subSec:DataExchange}) is set to 100ms. By default, the top 25\% target edge servers in the system with the highest available bandwidth are employed as entry edge servers to alleviate bandwidth bottlenecks (\cref{subSec:Reliable_AppVendorInteraction}). We vary seven setting parameters (summarized in Table \ref{Tab:ExperimentParameters}) to evaluate EdgeDis on the testbed (\Cref{subsubSec:testbed}). Each time we vary one parameter while keeping others as default and report the averaged results over 200 experiment runs.

\begin{table}[!tb]
    \centering
    \renewcommand{\arraystretch}{1.2}
    \tabcaption{Parameter Settings}
    \vspace{-1em}
    \label{Tab:ExperimentParameters}
    \scalebox{0.9}{\begin{tabular}{l|cc}
        \hline
        \hline
        \textbf{Parameter} &   \textbf{Value}   & \textbf{Default}\\
        \hline
        Replica Scale ($n$) & 8, 16, 32, 64, 128 & 32 \\
        Network Density ($nd$) & 1.0, 1.2, 1.4, 1.6, 1.8, 2.0 & 1.4 \\
        Failure Rate ($r$) & 0\%, 0.2\%, 0.4\%, 0.6\%, 0.8\%, 1.0\% & 0\% \\
        Data Size ($ds$) & 256MB, 512MB, 1GB, 2GB, 4GB & 1GB \\
        Data Block Size ($bs$) & 128KB, 256KB, 512KB, 1MB, 2MB & 512KB \\
        Network Delay ($dl$) & [5, 15], [10, 25],[15, 40] & [5,15]\\            
        Cost Ratio ($cr$) & 5, 10, 20, 30, 40 & 20\\   
        \hline
        \hline
        \end{tabular}}
\end{table}

\begin{itemize}
    \item \textbf{Replica Scale ($n$)} - the number of replicas of $d$ to be disseminated, varies exponentially from $8$ to $128$ and is 32 by default. 
    
    \item \textbf{Network Density ($nd$)} - the ratio of the number of physical connections between all edge servers over the number of edge servers in the system. It varies from 1 to 2 in steps of 0.2, and is 1.4 by default. 

    \item \textbf{Failure Rate ($r$)} - the probability that a failure occurs during data transmission, it varies from 0\%, i.e., no failure, to 1\% in step of 0.2\%. For example, when $r = 0.2\%$, there is a 0.2\% probability that a data block transmission will fail. 
    
    \item \textbf{Data Size ($ds$)} - the size of data $d$ to be disseminated. It varies exponentially from 256MB to 4GB, and is 1GB by default.
    
    \item \textbf{Data Block Size ($bs$)} - the data block size of $d$. It varies exponentially from 128KB to 2MB, and is 512KB by default.
    
    \item \textbf{Network Delay ($dl$)} - the range of communication delay between two linked edge servers. The minimum delay varies from 5ms to 15ms while the maximum delay varies from 15ms to 40ms. 
    
    \item \textbf{Cost Ratio ($cr$)} - the ratio of cloud-to-edge transmission cost to edge-to-edge transmission cost when transmitting one unit of data. A greater value indicates that the transmission cost between edge servers is cheaper than the transmission cost between cloud and individual edge servers. $cr$ varies from 5 to 40 and is 20 by default.
\end{itemize}

\subsubsection{Competing Schemes}
\label{subsubSec:Baseline}

EdgeDis is compared against four representative schemes, including two baseline schemes, i.e., DataSync, Gossip, Raft, and a state-of-the-art scheme, i.e., EDD-A.

\begin{itemize}
    \item \textbf{DataSync}~\cite{datasync2018AWS} - In the experiments, multiple DataSync instances are used to transmit data $d$ to target edge servers concurrently. 
    \item \textbf{Gossip}~\cite{boyd2006randomized} - In the experiments, every data block of $d$ is transmitted from cloud server to one edge server in the system. Then, the edge server transmits replicas of the data block to all its neighbor edge servers, which continue to forward the replicas to their neighbors until all target edge servers receive a copy of the data block. 
    
     \item \textbf{Raft}~\cite{ongaro2014Raft} - A leader is randomly elected to coordinate the EDD process, as illustrated in Fig. \ref{fig:motivation}(d). The leader receives data blocks from the cloud and forwards them to target edge servers.   
     
    \item \textbf{EDD-A}~\cite{xia2021cost} - Cloud server transmits data blocks of $d$ to target edge servers along the paths in a Steiner tree (\cref{Sec:Introduction}).

\end{itemize}

\subsubsection{Performance Metrics}
\label{subsubSec:PerformanceMetrics}
As introduced in Section \ref{Sec:Background_Motivation}, we have three objectives to achieve when designing EdgeDis, including G1 Efficiency, G2 Economy, and G3 Reliability. To evaluate EdgeDis's ability to fulfill all three objectives, the following two metrics are employed for experiments. 

\begin{itemize}
    \item \textbf{Dissemination Time}~\cite{xia2021cost,xia2022formulating} - measured by the time taken to complete the EDD process. A short dissemination time serves G1 Efficiency.
    \item \textbf{Dissemination Cost}~\cite{xia2021cost,xia2022formulating}- the economic costs incurred by EDD process. To allow easy conversion to real-world monetary costs, we set transmitting data $d$ between two edge servers costs 1 unit, and set transmitting $d$ over the backhaul network costs according to $cr$. For example, when $cr=20$, transmitting $d$ over the backhaul network costs 20 units. A low cost serves G2 Economy.
\end{itemize}

Please note that the objective \textbf{G3 Reliability} can be measured by EdgeDis's efficiency and costs upon dissemination failures. Therefore, the reliability is not measured individually, and alternatively, it will be measured based on the above two metrics when the failure rate $r$ varies.

\subsubsection{Testbed}
\label{subsubSec:testbed}
A testbed is built with one cloud server and 128 edge servers. A c4.2xlarge EC2 VM with 8 vCPU and 16GB RAM is deployed on Amazon AWS as the cloud server, running Ubuntu 16.04. Its network bandwidth is 1Gbps, which ensures that the EDD process is not throttled by the cloud server during the experiment. A total of 128 virtual machines (VMs) are deployed in a local data center as edge servers, each equipped with 1 vCPU, 2GB RAM, and 1Gbps bandwidth, running Ubuntu 16.04. They are connected according to a given network density $nd$ (\cref{subsubSec:Experimental_Settings}) to build an edge caching system. The network topology between edge servers is a regular graph with a density of $nd$. Network tests indicate that the round-trip time taken to transmit a 512KB data block from the cloud server to the edge servers is 240.95ms on average.  EdgeDis is a generic data dissemination scheme, so synthetic data is employed to simulate the data to be disseminated.

A prototype of EdgeDis is implemented in Java 16 for evaluation. It consists of two main modules, the cloud module and the edge module. The former runs on the cloud server for generating and transmitting data blocks to the edge caching system. An edge module runs on each of the edge servers in the system for receiving and transmitting data blocks. The algorithms used in EdgeDis are presented by Pseudocode 1 to Pseudocode 6 in Appendix A.

\subsection{Experimental Results}
\label{subSec:Experimets_ExperimentResults}

\subsubsection{Overall Performance (\textit{n = 32, nd = 1.4, ds = 1}GB, \textit{bs = 512}KB)}
It can be easily observed from \textbf{Table \ref{tab:results}} that EdgeDis achieves the best performance in every case, in particular, with an order of gratitude advantage in time. When $r=0.0\%$, i.e., no data dissemination failure, EdgeDis reduces the dissemination time consumption by 85.64\%, 88.28\%, 86.36\%, and 88.27\%,  respectively, compared with DataSync, Gossip, Raft, and EDD-A. This shows EdgeDis's superior ability to achieve G1 Efficiency by 1) partitioning data $d$ into multiple data blocks for simultaneous transmission (\cref{Sec:Reliable_EDD}) and 2) transmitting data blocks to and within the edge caching system asynchronously \cref{Sec:Reliable_EDD}.

\setlength{\textfloatsep}{4pt}
\begin{table}[!tb]
    \centering
    \renewcommand{\arraystretch}{1.2}
    \tabcaption{Performance Comparison}
    \vspace{-1em}
    \label{tab:results}
    \centering
    \scalebox{0.85}{\begin{tabular}{c|c|c|c|c|c|c}
    \hline
    \hline
       \textbf{Metric} & \textbf{$r$} & \textbf{DataSync} & \textbf{Gossip} & \textbf{Raft} & \textbf{EDD-A} & \cellcolor{lightgray} \textbf{EdgeDis} \\
    \hline
       \multirow{2}{*}{Time (seconds)} & 0 & 562.9 & 689.7 & 592.5 & 688.8 & \cellcolor{lightgray} 80.8 \\
       & 0.6\% & 661.1 & 810.4 & 595.4 & 693.8& \cellcolor{lightgray} 81.2\\
    \hline
       \multirow{2}{*}{Cost} & 0 & 640.0  & 64.8  & 52.0  & 100.4 & \cellcolor{lightgray} 52.0  \\
       & 0.6\% &  643.8 & 65.1   & 52.3  & 117.9 & \cellcolor{lightgray} 52.3  \\
    \hline
    \hline
    \end{tabular}}
\end{table}

Table \ref{tab:results} also shows the ability of EdgeDis to achieve G2 Economy. Compared with DataSync and Edd-A, it costs an average of 91.87\% and 48.21\% less, respectively, to complete an EDD process. These advantages come from the low costs of transmitting data $d$ out of the cloud only once (\cref{subSec:Reliable_AppVendorInteraction}) with $r=0$. Gossip and Raft also take this advantage and incur the same backhaul network costs as EdgeDis. However, Gossip broadcasts data between all edge servers in the system, which incurs higher costs. Under EDD-A, the volume of data transmitted out of the cloud depends on the structure of the Steiner tree. It often connects the cloud server (the root node) to multiple entry edge servers. Thus, EDD-A incurs higher costs than EdgeDis and Raft. DataSync incurs the most costs because it has to transmit a replica of $d$ to each of the target edge servers directly from the cloud server (\cref{Sec:Introduction}).

When dissemination failures occur with $r=0.6\%$, all the four competing schemes take more time to complete an EDD process. This immediately translates to higher costs. However, when $r$ increases from 0 to 0.6\%, DataSync's and EDD-A’s time consumption increases by 17.46\%, having to transmit the entire data block from the cloud every time the transmission of a data block fails. In contrast, EdgeDis's time consumption increases by 0.49\% only, much less than DataSync and EDD-A. This illustrates EdgeDis's ability to achieve G3 Reliability. The reasons that EdgeDis is more resilient to data dissemination failures are twofold: 1) handling internal dissemination failures over the edge network instead of the backhaul network (\cref{subSec:DataExchange}); and 2) selecting a coordinator to supply missing data blocks within the system rather than from the remote cloud (\cref{subSec:MissingDataRecovery}).

 Now, we evaluate the impact of different parameters on EdgeDis's time consumption and cost, summarized in Table T1 and Table T2, respectively, in Appendix B.

\subsubsection{Performance vs. Replica Scale}

\begin{figure}
    \centering
    \subfigure[Time vs. \textit{n}]{
    \includegraphics[width=0.48\linewidth]{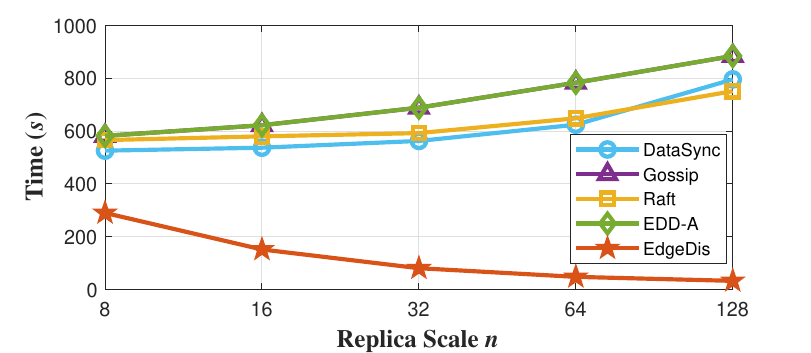}}
    \subfigure[Cost vs. \textit{n}]{
    \includegraphics[width=0.48\linewidth]{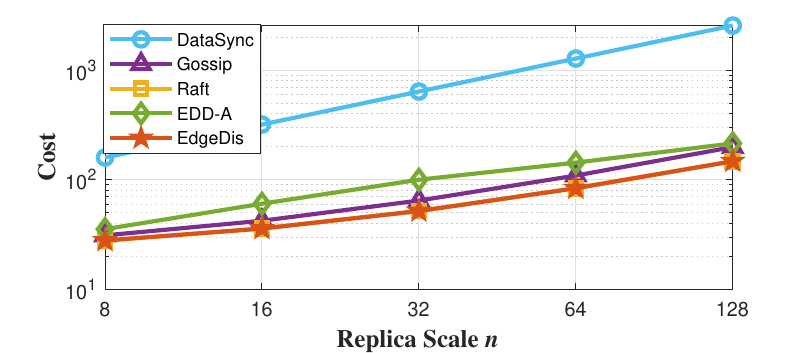}}
    \vspace{-1em}
    \caption{Impact of Replica Scale $n$}
    \vspace{-0.5em}
    \label{fig:EdgeServerScale}
\end{figure}

\textbf{Fig. \ref{fig:EdgeServerScale} }compares the performance of five competing schemes with different replica scales ($n$). It can be observed that EdgeDis achieves the best performance, in particular in terms of efficiency.

\textbf{Fig. \ref{fig:EdgeServerScale}(a)} demonstrates the average time taken by the five schemes to complete an EDD process when the replica scale $n$ increases from 8 to 128. An interesting observation is that when $n$ increases, DataSync, Gossip, Raft, and EDD-A take more time to complete an EDD process while EdgeDis takes less time. Take Raft as an example. The leader is solely responsible for transmitting data blocks to target edge servers. It becomes a performance bottleneck when there are a large number of target edge servers. Unlike Raft, EdgeDis disseminates data blocks through multiple entry edge servers in parallel - 25\% of the target edge servers in the experiments. A larger $n$ allows the cloud server to disseminate data blocks through more entry edge servers simultaneously. This increases the parallelism and significantly reduces the overall time. On average, EdgeDis takes 80.19\%, 80.84\%, 80.77\%, and 83.05\% less time to complete an EDD process compared with other schemes.

\textbf{Fig. \ref{fig:EdgeServerScale}(b)} shows that compared with DataSync, Gossip, and EDD-A, EdgeDis saves 92.98\%, 18.80\%, and 37.39\% of the transmission costs on average across all the cases. This indicates tremendous economic savings for app vendors. The figure also shows that the costs of all five schemes increase with the increase in $n$. This is not surprising because they all need to transmit more data overall to the target edge servers when $n$ increases. However, the cost incurred by DataSync increases by 15.00x when $n$ increases from 8 to 128. In contrast, the costs incurred by EDD-A, Raft, Gossip, and EdgeDis increase by 6.06x, 5.29x, 5.38x, and 5.28x, respectively, much less than DataSync. This demonstrates the advantage of shifting the workloads from the backhaul network to the edge network. DataSync always transmits data to each edge server directly from the cloud, and thus incurs the most extra backhaul network traffic when $n$ increases. EDD-A transmits data according to a Steiner tree with the cloud server as the root node and the target edge servers as leaf nodes (\cref{Sec:Introduction}). When $n$ increases, the cloud server has more child nodes in the Steiner tree. More traffic will be incurred over the backhaul network under EDD-A. In contrast, Raft transmits data from the cloud to only the leader, and EdgeDis transmits each data block to only one entry edge server. An increase in $n$ incurs extra traffic only over the edge network, which is much cheaper than transmission over the backhaul network.

\subsubsection{Reliability through Performance vs. Failure Rate} 

\begin{figure}
    \centering
    \subfigure[Time vs. \textit{r}]{
    \includegraphics[width=0.48\linewidth]{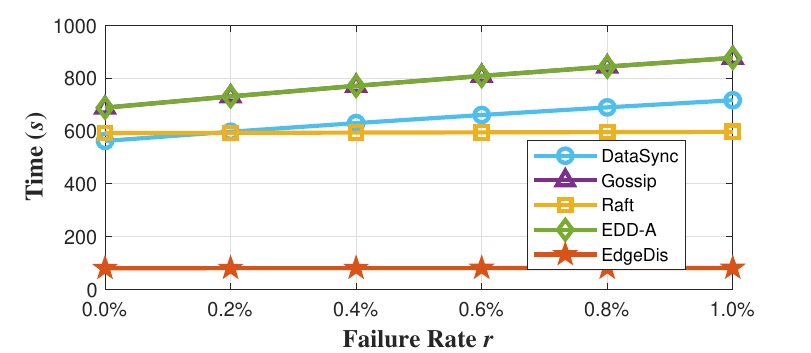}}
    \subfigure[Cost vs. \textit{r}]{
    \includegraphics[width=0.48\linewidth]{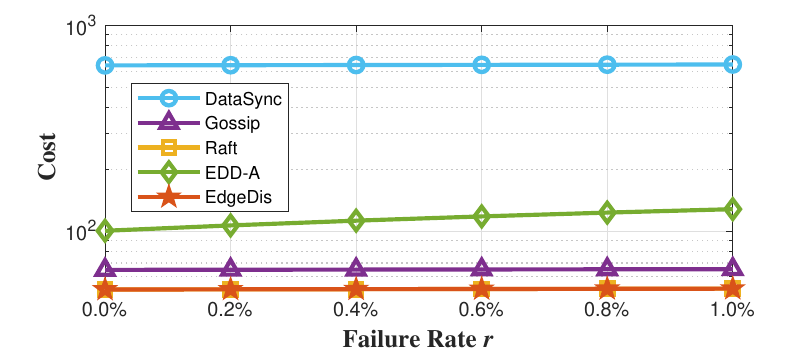}}
    \vspace{-1em}
    \caption{Impact of Failure Rate $r$}\label{fig:InterruptionRatio}
\end{figure}

As introduced in Section \ref{Sec:Introduction}, in the highly volatile MEC environment, data dissemination may be interrupted by various events. 
We examine EdgeDis's ability to adapt to dissemination failures in different EDD scenarios by varying the failure rate $r$ from 0.0\% to 1.0\%. \textbf{Fig. \ref{fig:InterruptionRatio}} shows EdgeDis's superior performance in all the cases.

We can see in \textbf{Fig. \ref{fig:InterruptionRatio}(a)} that with the increase in $r$, EdgeDis's dissemination time increases from 80.8s to 81.4s by only 0.74\%, much less significantly compared with DataSync, Gossip, Raft, and EDD-A's 27.40\%, 27.14\%, 0.83\%, and 27.39\% increases. When $r$ increases, dissemination failures are more likely to occur over the backhaul network and the edge network. Only in the former case will EdgeDis resend data blocks from the cloud. In the latter case, data blocks are re-transmitted over the edge network, which is much faster than over the backhaul network. If an entry edge server fails when only some of the target edge servers have received its data blocks, the coordinator will supply the remaining target edge servers with these data blocks (\cref{subSec:MissingDataRecovery}). This is done asynchronously with the dissemination of other data blocks and thus only delays EDD process slightly.

\textbf{Fig. \ref{fig:InterruptionRatio}(b)} shows that all five schemes incur higher costs when the failure rate $r$ increases. This is expected as they all need to re-transmit data upon dissemination failures. On average across all cases, EdgeDis's cost is 54.45\%, 19.67\%, and 91.87\% lower compared with DataSync, Gossip, and EDD-A. We can also see that EdgeDis's costs and extra costs caused by the increase in $n$ are almost the same as Raft. This evidences EdgeDis's comparable ability to adapt to dissemination failures compared with Raft. This is an exciting observation because it shows that EdgeDis does not sacrifice G3 Reliability for G1 Efficiency or G2 Economy upon dissemination failures.

\subsubsection{Performance vs. Network Density}

\begin{figure}
    \centering
    \subfigure[Time vs. \textit{nd}]{
    \includegraphics[width=0.48\linewidth]{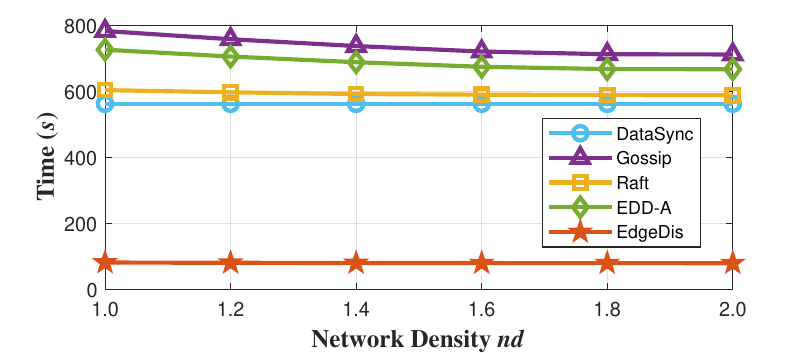}}
    \subfigure[Cost vs. \textit{nd}]{
    \includegraphics[width=0.48\linewidth]{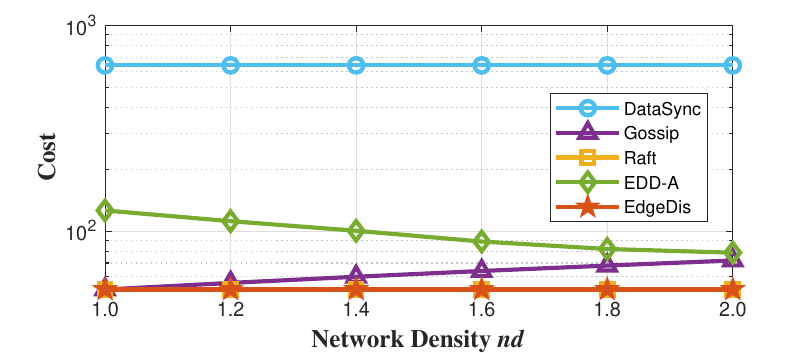}}
    \vspace{-1em}
    \caption{Impact of Network Density $nd$}
    \label{fig:NetworkDensity}
\end{figure}

To study the impact of system topology on EdgeDis's performance, we vary the network density $nd$ from 1 to 2 in steps of 0.2. As shown in \textbf{Fig. \ref{fig:NetworkDensity}}, EdgeDis again excels in all the cases.

As shown in \textbf{Fig. \ref{fig:NetworkDensity}(a)}, Gossip, EDD-A, Raft, and EdgeDis's dissemination times decrease slightly by 9.05\%, 8.15\%, 2.60\%, and 2.45\%, respectively, when $nd$ increases from 1.0 to 2.0. That is because when edge servers are more connected, their longest and average distances decrease and it takes less time to complete the part of EDD within the system, except under DataSync. DataSync's dissemination time does not react to the change in $nd$ because it does not transmit data over the edge network at all. \textbf{Fig. \ref{fig:NetworkDensity}(b)} shows that an increase in $nd$ reduces EDD-A's cost. With a larger $nd$, each edge server is linked to more edge servers in the system on average. This allows EDD-A to find a shorter Steiner tree for EDD which saves the costs. Interestingly, Gossip's cost increases when $nd$ is larger. That is because Gossip transmits data along all network connections.

\subsubsection{Performance vs. Data Size}

\begin{figure}[tbp]
\begin{minipage}[c]{0.49\linewidth}
    \centering
    \includegraphics[width=\linewidth]{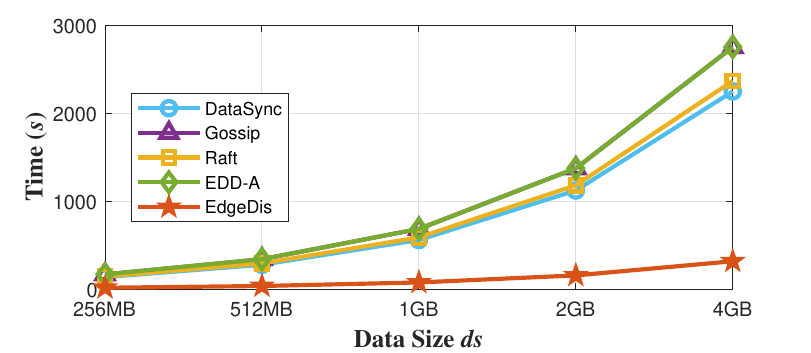}
    \caption{ Impact of Data Size $ds$} 
    \label{Fig:Efficiency_DataSize} 
\end{minipage}
\begin{minipage}[c]{0.49\linewidth}
    \centering 
    \includegraphics[width=\linewidth]{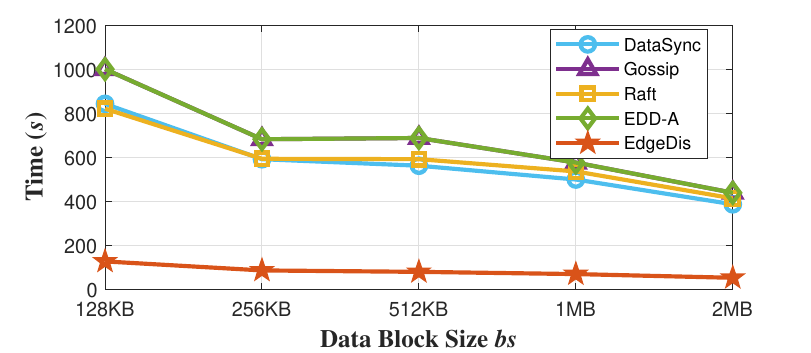} 
    \caption{Impact of Data Block Size $bs$ } 
    %
    \label{Fig:Efficiency_DataBlockSize} 
\end{minipage}
\end{figure}

\textbf{Fig. \ref{Fig:Efficiency_DataSize}} compares the time taken by the competing schemes to disseminate data of different sizes ($ds$). Note that the change in $ds$ impacts dissemination costs in almost the same way as dissemination time. Thus, their dissemination costs are omitted here. In the figure, we can see that all the schemes take more time to complete an EDD process when $ds$ increases. EdgeDis again manifests its significant advantages. When $ds$ increases from 256MB to 4GB by 16x, its dissemination time increases from 20.1s to 321.3s. In the meantime, DataSync, Gossip, Raft, and EDD-A's dissemination times increase from 140.7s, 171.1s, 148.1s, and 172.2s to 2251.4s, 2733.7, 2370.2s, and 2755.3s, respectively.

\subsubsection{Performance vs. Data Block Size} 

EdgeDis partitions a data item $d$ into multiple data blocks for dissemination to improve efficiency. To study how it impacts EdgeDis's performance, we vary the data block size $bs$ in the experiments. As shown in \textbf{Fig. \ref{Fig:Efficiency_DataBlockSize}}, when there is no dissemination failure, a larger $bs$ reduces the dissemination time for all the competing schemes. The reason is that given a data item of a fixed size (1GB in the experiments) and a larger $ds$, EdgeDis partitions $d$ into fewer data blocks for dissemination. This leads to fewer server interactions during the EDD process, which reduces the overall dissemination time. However, when a dissemination failure occurs, the impacted data block needs to be re-transmitted. This will incur extra costs. There is a time-cost trade-off. In practice, the data block size can be set empirically by the app vendor according to its time and cost needs.

\subsubsection{Performance vs. Cost Ratio}

EdgeDis achieves G2 Economy by reducing backhaul network traffic and forwarding data between edge servers through fronthaul edge server network, as the transmission cost on backhaul network is much higher than edge server network \cite{xia2021cost,li2020read}. Now we study the impact of cost ratio ($cr$) on EdgeDis' performance. Please note that $cr$ affects only the transmission cost, and thus the comparison of time consumption is omitted.

From Fig. \ref{fig:costRatio} we can observe that the overall cost of every scheme increases along with the increase in $cr$. This is straightforward, as a greater $cs$ means transforming a piece of data from the cloud is more expensive. Due to the same reason, EdgeDis's advantages over DataSync and EDD-A become greater when $cr$ increases. For example, it costs 76.88\% and 34.95\% less than DataSync and EDD-A, respectively, when $cr=5$. However, when $cr$ increases to 40, EdgeDis costs 94.37\% and 65.93\% less than DataSync and EDD-A, respectively. Same to EdgeDis, both Gossip and Raft transmit one data from the cloud to the system, thus the advantages of EdgeDis over Gossip and Raft do not change significantly.

\subsubsection{Performance vs. Network Delay}

\begin{figure}[tbp]
\begin{minipage}[c]{0.49\linewidth}
    \centering
    \includegraphics[width=\linewidth]{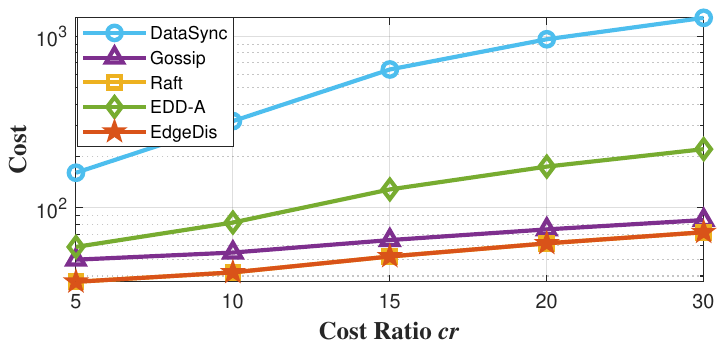}
    \vspace{-1em}
    \caption{Impact of Cost Ratio $cr$}
    \vspace{-0.5em}
    \label{fig:costRatio}
\end{minipage}
\begin{minipage}[c]{0.49\linewidth}
    \centering 
    \includegraphics[width=\linewidth]{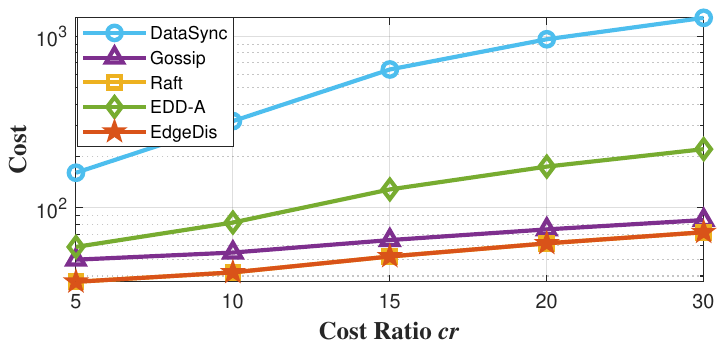}
    \vspace{-1em}
    \caption{Impact of Network Delay}
    \label{fig:networkdelay}
    \vspace{-0.5em}
\end{minipage}
\end{figure}

In practice, edge servers may be far from the others, which prolongs the network latency between them. Now, we study the impact of network delay $dl$ on EdgeDis's performance by varying $dl$ from [5,15] to [15,40]. Fig. \ref{fig:networkdelay} depicts the results. Briefly, when $dl$ increases, all schemes take more time to complete the dissemination. However, the time consumption of DataSync, Gossip, and EDD-A increase significantly. In contrast, the time consumption of Raft and EdgeDis increase slightly. The reason is that both Raft and EdgeDis have special mechanisms to tackle the slow edge servers. Taking EdgeDis as an example, an entry edge server commits a data block dissemination when it transmits the data block to majority of edge servers. Then, the cloud server can start to transmit new data blocks. At the same time, the entry edge server continues to transmit its data block to slow edge servers. Therefore, the higher network delays of a few edge servers do not significantly impact the overall efficiency of EdgeDis. This also evidences EdgeDis's superiority in achieving G1 Efficiency. 

\begin{figure}
    \centering
    \subfigure[Commu. Overheads vs. \textit{n}]{
    \includegraphics[width=0.48\linewidth]{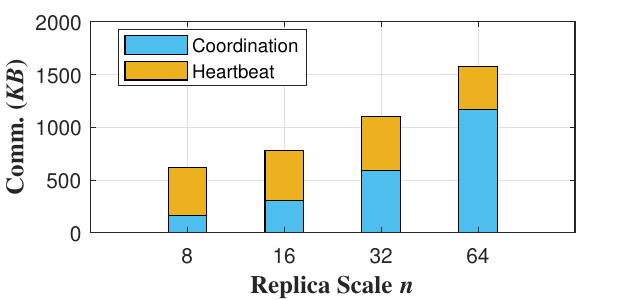}}
    \subfigure[Commu. Overheads vs. \textit{ds}]{
    \includegraphics[width=0.48\linewidth]{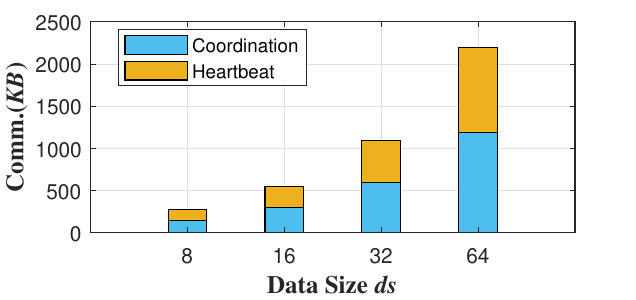}}
    \vspace{-1em}
    \caption{Extra Overheads incurred by EdgeDis}
    \label{fig:System_Overheads}
\end{figure}

\subsubsection{Extra Overheads Examination} 
\label{subsubsec:systemOverhead}
EdgeDis employs a distributed consensus to coordinate the EDD process which incurs extra communication overheads. First, there are extra \textit{coordination} overheads when servers exchange various messages to coordinate an EDD process (\cref{Sec:Reliable_EDD}), which include meta information like data block IDs, data ID, etc. Second, coordinator broadcasting \textit{heartbeats} also incurs extra communication overheads. These extra communication overheads are presented in \textbf{Fig. \ref{fig:System_Overheads}}.

In \textbf{Fig. \ref{fig:System_Overheads}(a)}, we can find that the extra communication overheads incurred by EdgeDis are low in relevance to the replica scale $n$, i.e., the number of receivers. For example, when $n = 8$, a total of 616KB communication overheads are incurred to coordinate the entire EDD process, about 77KB per receiver. When $n$ increases, the coordination becomes more complicated but EdgeDis's communication overheads increase only linearly, hitting 1,572.4KB when $n=64$. Interestingly, the communication overheads per receiver decrease to 24.57KB. We investigated and found the reason. A larger $n$ increases the number of entry edge servers, accelerating the EDD process with higher parallelism, as shown before in Fig. \ref{fig:EdgeServerScale}(a). Thus, fewer heartbeats are broadcasted during the entire process.

\textbf{Fig. \ref{fig:System_Overheads}(b)} illustrates EdgeDis's communication overheads incurred for disseminating data items of different sizes to 32 receivers. As demonstrated in Fig. \ref{Fig:Efficiency_DataSize}, EdgeDis takes more time to disseminate a larger data item partitioned into more data blocks. As a result, the corresponding coordination overheads and heartbeat overheads both increase accordingly.

\subsubsection{Impact of Entry Edge Server Selection Strategy} EdgeDis selects edge servers with the highest bandwidth as entry edge servers (\cref{subSec:Reliable_AppVendorInteraction}). To study the impact of this mechanism, we create a version of EdgeDis named EdgeDis-RND that randomly selects entry edge servers regardless of their bandwidths. In this experiment, we throttle edge servers' bandwidths randomly with DummyNet\footnote{http://info.iet.unipi.it/\textasciitilde luigi/dummynet/} to simulate their bandwidth heterogeneity.

\begin{figure}
    \centering
    \subfigure[Impact of Entry Edge Server Amount]{
    \includegraphics[width=0.48\linewidth]{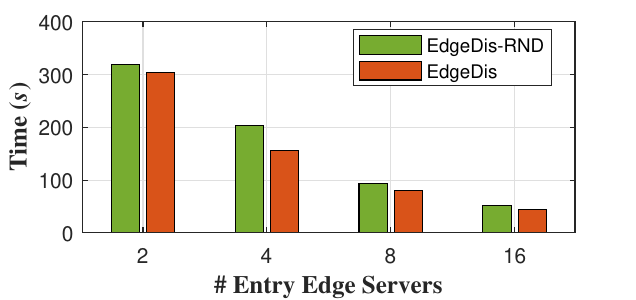}}
    \subfigure[Impact of Bandwidth Heterogeneity]{
    \includegraphics[width=0.48\linewidth]{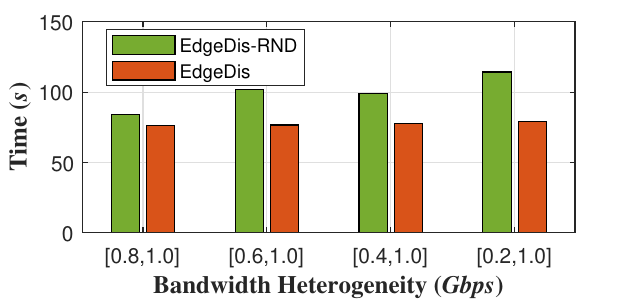}}
    \vspace{-1em}
    \caption{Effect of Entry Server Selection}
    \label{fig:Impact_Entry_Selection}
\end{figure}

\textbf{Fig. \ref{fig:Impact_Entry_Selection}(a)} compares their dissemination times for disseminating 1GB of data to 32 receivers when edge servers' bandwidths range between 0.6Gbps and 1.0Gbps. It is not hard to observe EdgeDis's significant advantages in any case. On average, EdgeDis takes 16.33\% less time than EdgeDis-RND to complete an EDD process. An interesting finding is that EdgeDis's advantage decreases from 22.86\% to 8.36\% when the number of entry edge servers increases from 4 to 16. The reason is that when more entry edge servers are selected from receivers, those new entry edge servers are more likely to become a performance bottleneck due to their lower average bandwidth. This reduces the performance gap between EdgeDis and EdgeDis-RND. 

\textbf{Fig. \ref{fig:Impact_Entry_Selection}(b)} compares their dissemination times with different levels of bandwidth heterogeneity, with 8 entry edge servers, $ds = 1GB$ in a system comprised of 32 edge servers. We can find that when edge servers become more heterogeneous, EdgeDis gains more advantages over EdgeDis-RND, i.e., 7.16\% to 27.33\%. This further evidences the usefulness of selecting entry edge servers in the highly heterogeneous MEC environment.

\subsubsection{Efficiency of Coordinator Election/Re-election}
\label{subsubsec:parameterImpacts}

Now we investigate EdgeDis's efficiency in coordinator election (\cref{Sec:CoordinatorElection}), measured by the time EdgeDis takes to elect a new coordinator when the system starts. This also indicates EdgeDis's efficiency in handling coordinator failures (\cref{subSec:MissingDataRecovery}).

\begin{figure}[tbp]
\centering
    \centering
    \includegraphics[width=0.5\linewidth]{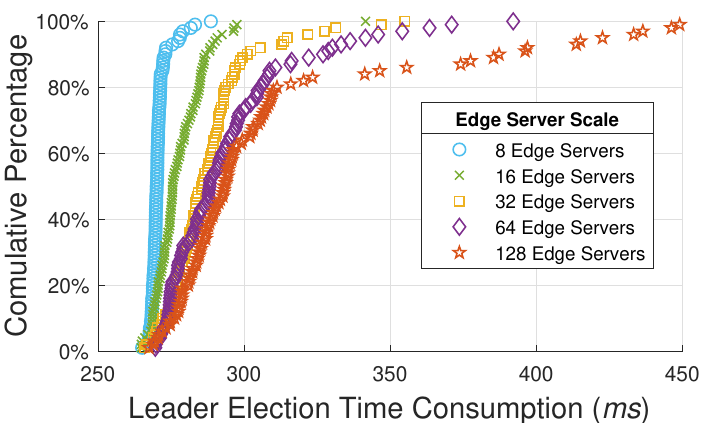}
    \vspace{-1em}
    \caption{Election Efficiency vs. $n$}
    \label{Fig:Election_ServerScale} 
\end{figure}

\smallskip
\textbf{Impact of Replica Scale $n$.} The cumulative distribution function (CDF) plot shown in \textbf{Fig. \ref{Fig:Election_ServerScale}} reports EdgeDis's coordinator election/re-election time when replica scale $n$ increases from 8 to 128 with coordinator timeout $t$ set at 250ms. We can see that EdgeDis needs more time to elect a new coordinator when there are more edge servers in the system. The reason is that when $n$ increases, candidate edge servers have to send vote requests to more edge servers to seek their support. This immediately increases the overall time needed to complete the coordinator election. The increase in $n$ also raises the probability of vote splitting, i.e., no competing candidate edge server can obtain majority support in the system. In such cases, coordinator election iterates and takes more time to complete eventually. This can be clearly seen in Fig. \ref{Fig:Election_ServerScale}. When $n=8$, coordinator election takes 264.95ms to 288.48ms to complete. When $n$ increases to 128, it takes 267.31ms to 454.08ms to complete.

\begin{figure}[tbp]
    \centering 
    \includegraphics[width=0.5\linewidth]{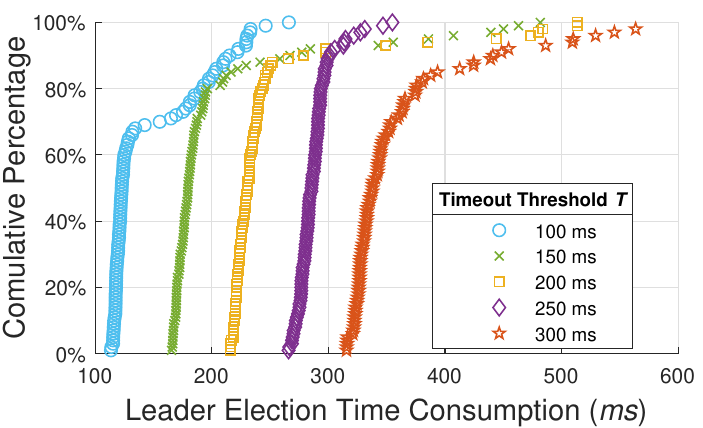} 
    \vspace{-1em}
    \caption{Election Efficiency vs. $t$}
    \label{Fig:Election_Timeout}  
\end{figure}

\smallskip
\textbf{Impact of Coordinator Timeout $t$.} Coordinator timeout $t$ (see Section \ref{Sec:CoordinatorElection}) determines when an edge server starts the coordinator election/re-election process. \textbf{Fig. \ref{Fig:Election_Timeout}} depicts the overall time taken to complete the election where $t$ varies from 100ms to 300ms in steps of 50ms, with $n=32$. When $t$ is small, edge servers start their own election processes soon when the coordinator fails. As a result, EdgeDis can elect a new coordinator rapidly. However, a smaller $t$ increases the number of candidates contending for the coordinator role at the same time. If a candidate cannot win the election rapidly, the other edge servers may start their own election processes. Intense competition will increase the probability that no candidate edge servers can obtain majority support from the edge servers in the system within a short time period. In such cases, it takes more time to elect a new coordinator, similar to when $n$ increases as shown in Fig. \ref{Fig:Election_ServerScale}. When $t$ increases, the competition becomes less intense and it is easier to elect a new coordinator (with fewer iterations). However, when $t$ is overly large, edge servers may wait too long before starting their election processes, which causes a delay in data transmission. This indicates that an optimal $t$ needs to be experimentally obtained to achieve the best performance for a specific system in practice.

\subsubsection{Coordinator Uniqueness}

\begin{figure*}[tbh]
    \centering
    \includegraphics[width=0.8\linewidth]{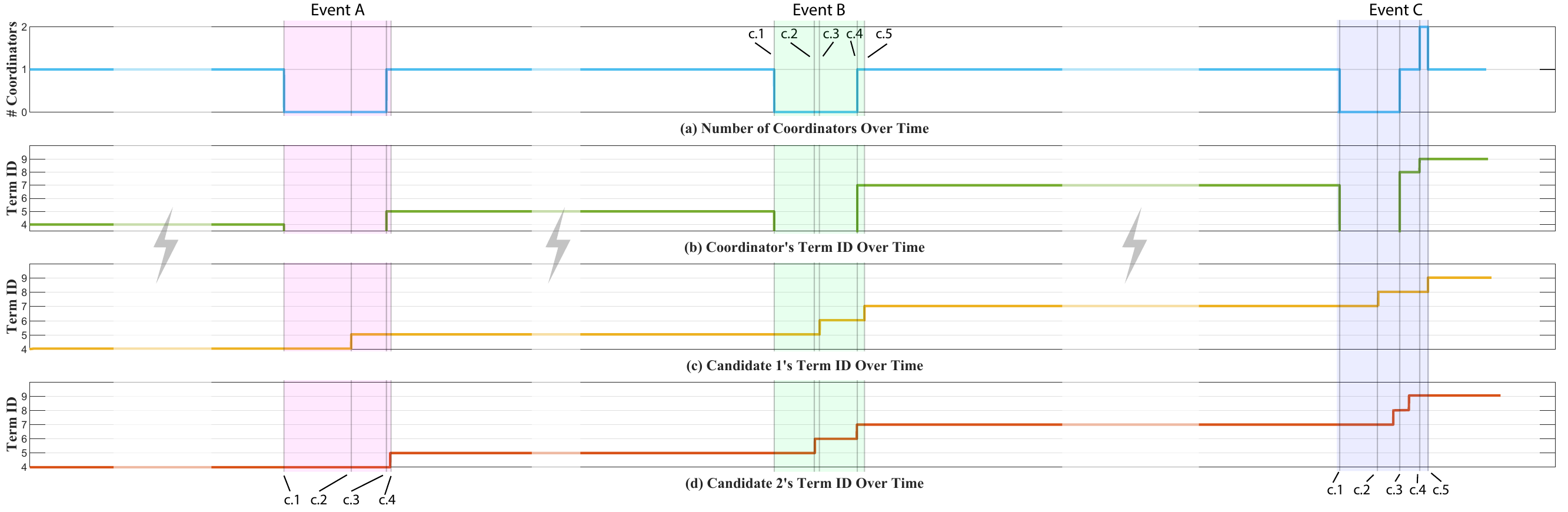}
    \vspace{-1em}
    \caption{Election Process to Ensure Coordinator Uniqueness. Two Candidates (1\&2) are Contending for Leadership.}
    \vspace{-1em}
    \label{fig:coordinatorUniqueness}
\end{figure*}

Now we experimentally study EdgeDis's ability to ensure the coordinator uniqueness. Fig. \ref{fig:coordinatorUniqueness} presents the results, in which we report 1) the total number of coordinators in the system, 2) the coordinator's term ID (0 if no coordinator), 3) the term ID of Candidate 1, and 4) the term ID of Candidate 2. We omit other follower edge servers for simplification. The initial coordinator has a term ID of 4. Then, we activate three events, denoted as A, B, and C, respectively, by manually killing the current coordinator.

In Event A, the coordinator with term ID of 4 fails at time $a.1$. Suppose  Candidate 1 increases its term ID and starts the election process at time $a.2$. It receives support from the majority of edge servers at $a.3$. Then, it starts to send out heartbeats to all other edge servers. When Candidate 2 receives the heartbeat and recognizes the existence of a new coordinator at time $a.4$, it becomes a follower and updates its term ID to 5, which is the term ID of Candidate 1.
 
In Event B, the current coordinator (Candidate 1) fails at time $b.1$. When timeout at $b.2$, Candidate 2 increases its term ID by 1 to 6 and starts the election process. Later at $b.3$, Candidate 1 increases its term ID by 1 to 6 and starts the election process. As both of them have the same term ID, the vote split occurs. In the next attempt, Candidate 2 increases its term ID by 1 to 7. It successfully receives support from the majority of edge servers at time $b.4$. Similar to the phenomenon in Event A, Candidate 1 becomes a follower at time $b.5$ and updates its term ID to 7. 

In the above two scenarios, there is at most one coordinator in the system. Now, we produce a specific scenario (Event C) in which there is more than one coordinator in the system. We kill Candidate 2 (current coordinator) at time $c.1$. Then, Candidate 1 contends for the leadership at time $c.2$ with term ID of 8 and becomes coordinator at $c.3$. During this time, Candidate 2 updates its term ID to 8 and starts its election process. Unfortunately, as most edge servers have supported Candidate 1, Candidate 2 cannot receive enough support with term ID of 8. Due to the network issue (we manually blocked the heartbeat sent by Candidate 1), Candidate 1 does not send its heartbeat to Candidate 2. Therefore, Candidate 2 increases its term ID to 9 and attempts again. As it has a greater term ID, it receives majority support at time $c.4$ and becomes a coordinator. In this case, both Candidate 1 and Candidate 2 are coordinators in the system. However, Candidate 1 will become a follower when it receives heartbeats from Candidate 2 or heartbeat responses from other edge servers (Lines 15-17 in Appendix A Pseudocode 3) at $c.5$, as its term ID is smaller. Thus, even if multiple coordinators can be elected in a very short time period, e.g., less than 50 ms, only one can survive.



\vspace{-1em}
\section{Conclusion and Future Work}
\label{Sec:Conclusion}

In this paper, we studied the edge data dissemination problem and proposed an innovative scheme named EdgeDis by distributed consensus to facilitate data dissemination from the cloud to edge caching systems. Under EdgeDis, multiple edge servers are involved in the data dissemination process to ensure high data distribution efficiency. Each data is sent from the cloud to only one edge server in the system to reduce the backhaul network traffic, i.e., to ensure high economy. A coordinator is elected to supply missing data blocks to edge servers to improve reliability. Experimental results demonstrate EdgeDis's superior performance at the price of slight system overheads. 

In the future, we will investigate the possibility of allowing EdgeDis to resist malicious nodes and cyberattacks. We will explore if coded settings and/or other recovery mechanisms can be employed by EdgeDis to further improve its performance. We will explore if incorporating multiple coordinators in the system can further boost the dissemination efficiency. We also will study the measures of edge server capacities to help choose suitable entry edge servers. Finally, we will design an optimized algorithm to find the most suitable number of entry edge servers introduced in Section \ref{subSec:Reliable_AppVendorInteraction}.


\bibliographystyle{IEEEtran}
\bibliography{bibfile}

\vskip -2.6\baselineskip plus -1fil
\begin{IEEEbiography}[{\includegraphics[width=1in,height=1.25in,clip,keepaspectratio]{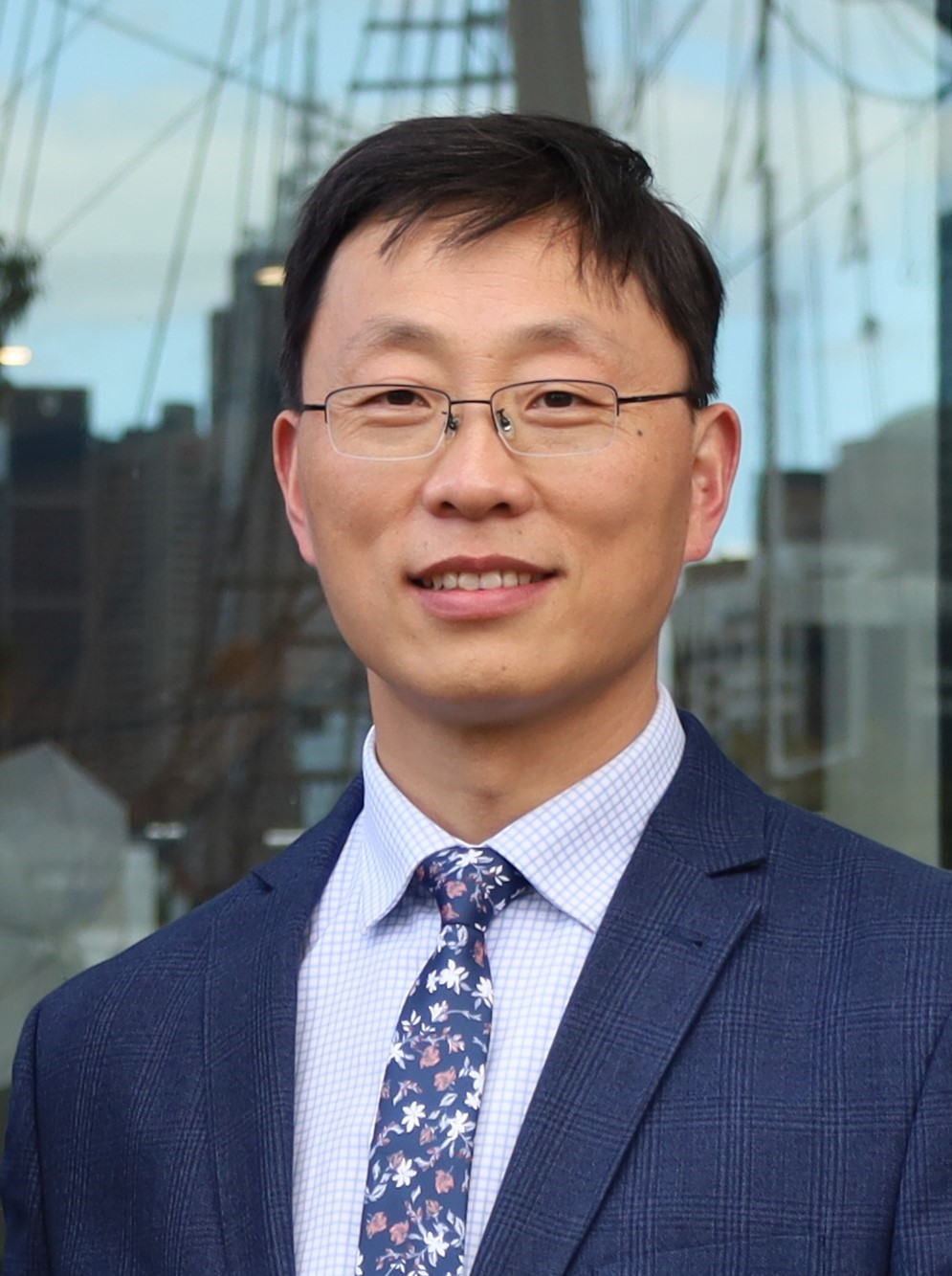}}]{Bo Li} received the BS and MS degree from the School of Information Science and Technology from Shandong Normal University in 2003 and 2010, respectively. He received his PhD degree from Swinburne University of Technology in 2023. He worked as an academic visitor at Shandong University from 2014 to 2015 and at Swinburne University of Technology in 2018. He is currently a lecturer at Victoria University. His research interests include software engineering, edge computing, and network security issues.
\end{IEEEbiography}

\vskip -2.6\baselineskip plus -1fil
\begin{IEEEbiography}[{\includegraphics[width=1in,height=1.25in,clip,keepaspectratio]{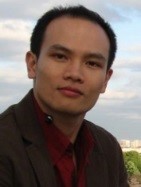}}]{Qiang He} received his first PhD degree from Swinburne University of Technology, Australia, in 2009 and his second PhD degree in computer science and engineering from Huazhong University of Science and Technology, China, in 2010. He is a Professor at Swinburne. His research interests include mobile edge computing, software engineering, service computing and cloud computing. More details about his research can be found at https://sites.google.com/site/heqiang/.
\end{IEEEbiography}

\vskip -2.6\baselineskip plus -1fil
\begin{IEEEbiography}[{\includegraphics[width=1in,height=1.25in,clip,keepaspectratio]{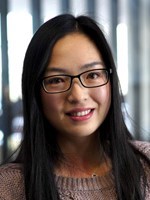}}]{Feifei Chen} received her PhD degree from Swinburne University of Technology, Australia in 2015. She is a lecturer at Deakin University. Her research interests include software engineering, cloud computing and green computing.
\end{IEEEbiography}

\vskip -2.6\baselineskip plus -1fil
\begin{IEEEbiography}[{\includegraphics[width=1in,height=1.25in,clip,keepaspectratio]{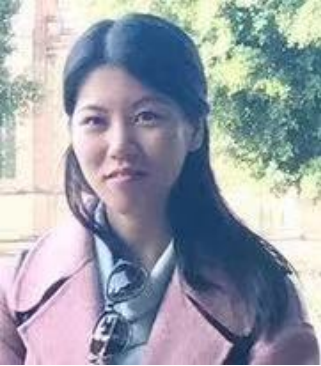}}]{Lingjuan Lyu} is a senior research scientist and Privacy-Preserving Machine Learning (PPML) team leader in Sony AI. She received Ph.D. from the University of Melbourne. She was a winner of the IBM Ph.D. Fellowship Worldwide. Her current research interest is trustworthy AI. She had published over 50 papers in top conferences and journals, including NeurIPS, ICML, ICLR, Nature, AAAI, IJCAI, etc. Her papers had won several best paper awards and oral presentations from top conferences.
\end{IEEEbiography}

\vskip -2.6\baselineskip plus -1fil
\begin{IEEEbiography}[{\includegraphics[width=1in,height=1.25in,clip,keepaspectratio]{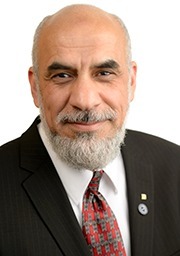}}]{Athman Bouguettaya} (Fellow, IEEE) received the PhD degree in computer science from the University of Colorado Boulder, USA, in 1992. He
is currently a professor with the School of Computer Science, University of Sydney, Australia. He is or has been on the editorial boards of several
journals including, the IEEE Transactions on Services Computing, ACM Transactions on Internet Technology, the International Journal on Next Generation Computing, and VLDB Journal. He was the recipient of several federally competitive grants in Australia, including ARC and in the U.S., including NSF and NIH. He is a distinguished scientist of the ACM.
\end{IEEEbiography}

\vskip -2.6\baselineskip plus -1fil
\begin{IEEEbiography}[{\includegraphics[width=1in,height=1.25in,clip,keepaspectratio]{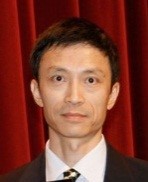}}]{Yun Yang} received his PhD degree from the University of Queensland, Australia, in 1992, in computer science. He is currently a full professor in the School of Science, Computing and Engineering Technologies at Swinburne University of Technology, Melbourne, Australia. His research interests include software technologies, cloud and edge computing, workflow systems, and service computing.
\end{IEEEbiography}

\clearpage

\renewcommand{\thesection}{\Alph{section}.\arabic{section}}
\setcounter{section}{0}
\begin{appendices}

\section{Algorithms Used in EdgeDis}
\numberwithin{equation}{section}
\label{Appendix:ApproachDesign_DataStructure}

\balance

\setlength{\textfloatsep}{4pt}
\begin{algorithm}[tbh]\small
\caption{\footnotesize{\textbf{Sender}'s Data Block Transmission Workflow}}
\label{pseudo_sender}
    \SetKwProg{Upon}{Upon}{ do}{}
    \DontPrintSemicolon
    \tcp{\colorbox{blue}{\textcolor{white}{Stage 2 Step 1:}} data block transmission message generation}
    \Upon{receipt of data block $d_i$ from $\mathbi{cs}$}
    {
       creates a data block transmission message $mes_s$\;
       $mes_s.dataBlock = d_i$    \tcp*{attach data block $d_i$}
       $mes_s.dataBlockId = i$\;
       sends $mes_s$ to all the other edge servers\;
    }
    \tcp{\colorbox{blue}{\textcolor{white}{Stage 2 Step 3:}} progress inspection}
    \Upon{receipt of a data block receipt message}
    {
        \If{received $\lceil (n+1)/2\rceil-1$ such messages for $d_i$}
        {
            responds $\mathbi{cs}$ a data block distribution completion message for $d_i$
        }
    }
\end{algorithm}
\vspace{-1em}

\setlength{\textfloatsep}{10pt}
\begin{algorithm}[tbh]\small
\caption{\footnotesize{\textbf{Receiver}'s Data Block Transmission Workflow}}
\label{pseudo_receiver}
    \SetKwProg{Upon}{Upon}{ do}{}
    \DontPrintSemicolon
    \tcp{\colorbox{blue}{\textcolor{white}{Stage 2 Step 2:}} data block receipt message generation}
    \Upon{receipt of message $mes_s$ from sender $s_j$}
    {
       receives and stores data block $d_i$ from $mes_s$\;
       $s_k.dataBlockStatus[i] = 1$ \tcp*{updates status}
       \If{$s_k.maxBlockId<i$} 
       {
        sets $s_k.maxBlockId = i$  \tcp*{$s_j$'s maximum data block ID}
       }
       generates data block receipt message $mes_r$\;
       $mes_r.dataBlockId=i$\;
       $mes_r.status = true$\;
       sends $mes_r$ to $s_j$\;
     }
\end{algorithm}

\setlength{\textfloatsep}{4pt}
\begin{algorithm}[bh]\small
\caption{\footnotesize{\textbf{Coordinator} $\mathbi{c}$}'s Data Block Collection Workflow}
\label{pseudo_dataCollection_coordinator}
    \SetKwProg{Upon}{Upon}{ do}{}
    \SetKwFor{Loop}{Loop}{}{}
    \DontPrintSemicolon
    \tcp{\colorbox{blue}{\textcolor{white}{Stage 3 Step 1-1}} reuse heartbeats}
    \Loop{}
    {
        creates a new heartbeat message $mes_h$ \tcp*{same to~\cite{ongaro2014Raft}}
        $mes_h.maxBlockId = \mathbi{c}.maxBlockId$ \tcp*{$\mathbi{c}$ perceived blocks}
        $mes_h.currentTermId=\mathbi{c}.termId$ \tcp*{maintains leadership}
        $mes_h.coordinatorId = \mathbi{c}.Id$  \tcp*{maintains leadership}
        broadcasts $mes_h$;
    }
    \Upon{receipt of heartbeat receipt $mes_{hr}$ from $\mathbi{f}$}
    {   \tcp{\colorbox{blue}{\textcolor{white}{Stage 3 Step 1-2}} perceive new data blocks}
       \tcp{new data blocks exist}
       \If{$mes_{hr}.maxBlockId > c.maxBlockId$}{
           updates $\mathbi{c}.dataBlockStatus[]$\;
           $\mathbi{c}.maxBlockId = mes_{hr}.maxBlockId$\;
       }
       checks $\mathbi{f}$'s data block possession status\;
       updates $\mathbi{c}.disseStatus[f.id][]$\;
       \tcp{\colorbox{blue}{\textcolor{white}{Stage 3 Step 1-3}} request missing data blocks}
       \If{missing data blocks are possessed by $\mathbi{f}$}
       {
            sends data block request to $\mathbi{f}$
       }
       \tcp{new coordinator exists}
       \If{$mes_{hr}.currentTermId > \mathbi{c}.termId$}{
            $\mathbi{c}.state=follower$\;
            $\mathbi{c}.termId = mes_{hr}.currentTermId$ \tcp*{update term}
       }
     }
     \tcp{\colorbox{blue}{\textcolor{white}{Stage 3 Step 1-3}} receive data blocks}
     \Upon{receipt of a data block response}{
        stores the data blocks in $dataBlock[]$\;
        updates $\mathbi{c}.dataBlockStatus[]$ via $dataBlockId[]$
     }
\end{algorithm}

\newpage

\newpage

\setlength{\textfloatsep}{4pt}
\begin{algorithm}[tbh]\small
\caption{\footnotesize{\textbf{Follower} $\mathbi{f}$}'s Data Block Collection Workflow}
\label{pseudo_dataCollection_follower}
    \SetKwProg{Upon}{Upon}{ do}{}
    \SetKwFor{Loop}{Loop}{}{}
    \DontPrintSemicolon
    \tcp{\colorbox{blue}{\textcolor{white}{Stage 3 Step 1-2}} notify new data blocks}
    \Upon{receipt of heartbeat message $mes_{h}$}
    {   
       \tcp{new data blocks exist in the system, for \colorbox{blue}{\textcolor{white}{\tiny{Stage 3 Step 2}}}}
       \If{$mes_{h}.maxBlockId > f.maxBlockId$}{
           updates $\mathbi{f}.dataBlockStatus[]$\;
           $\mathbi{f}.maxBlockId = mes_{h}.maxBlockId$\;
       }
       creates heartbeat receipt message $mes_{hr}$\;
       $mes_{hr}.maxBlockId = \mathbi{f}.maxBlockId$\;
       initializes $mes_{hr}.missBlockId[]$ accordingly\;
       \tcp{check leadership}
       \If{$mes_{h}.currentTermId<f.termId$}{
            \tcp{notifies $\mathbi{c}$ the existance of new coordinator}$mes_{hr}.currentTermId = \mathbi{f}.termId$ 
       }
       \Else{ \tcp{$\mathbi{c}$ is new, follows $\mathbi{c}$}
            $\mathbi{f}.termId = mes_{hr}.currentTermId$\; 
            $\mathbi{f}.coordinatorId = mes_{hr}.coordinatorId$\; 
       }
       sends $mes_{hr}$ to $\mathbi{c}$\;
     }
     \tcp{\colorbox{blue}{\textcolor{white}{Stage 3 Step 1-3}} send missing data blocks}
     \Upon{receipt of a data block request from $c$}{
        sends required data blocks via a data block response\;
     }
\end{algorithm}

\setlength{\textfloatsep}{3pt}{
\begin{algorithm}[tbh]\small
\caption{\footnotesize{\textbf{Candidate} $\mathbi{c}_f$}'s Coordinator Election Workflow}
\label{pseudo_election_candidate}
    \SetKwProg{Upon}{Upon}{ do}{}
    \SetKwFor{Loop}{Loop}{}{}
    \DontPrintSemicolon
    \tcp{\colorbox{blue}{\textcolor{white}{Step 1}} vote request generation}
    \Upon{coordinator timeout}
    { 
        $\mathbi{c}_f.status = candidate$ \tcp*{becomes candidate}
        $\mathbi{c}_f.termId++$ \tcp*{updates to new term ID}
        creates vote request $req$\;
        $req.currentTermId = \mathbi{c}_f.termId$\;
        $req.candidateId = \mathbi{c}_f.id$\;
        $req.totalBlocks=$ total number of data blocks $\mathbi{c}_f$ has\;
        $\mathbi{c}_f.supportedId=\mathbi{c}_f.id$ \tcp*{supports itself}
        broadcasts $req$ to contend for leadership\;
     }
     \tcp{\colorbox{blue}{\textcolor{white}{Step 3}} vote counting}
     \Upon{receipt of a vote response $res$}{
        \tcp{succeeds, becomes new coordinator}
        \If{received ($\lceil (n+1)/2\rceil-1$)th responses with $supported = true$}{
            
            $\mathbi{c}_f.status = coordinator$\;
            starts to send out heartbeats (see Section \ref{subSec:MissingDataRecovery})
        }
        \tcp{other new coordinator exists, quits current election}
        \If{$res.currentTermId > \mathbi{c}_f.termId$}{
            $\mathbi{c}_f.status = follower$
        }
     }
\end{algorithm}}

\setlength{\textfloatsep}{3pt}
\begin{algorithm}[bh]\small
\caption{\footnotesize{\textbf{Follower} $\mathbi{f}$}'s Coordinator Election Workflow}
\label{pseudo_election_follower}
    \SetKwProg{Upon}{Upon}{ do}{}
    \SetKwFor{Loop}{Loop}{}{}
    \DontPrintSemicolon
    \tcp{\colorbox{blue}{\textcolor{white}{Step 2}} vote response generation}
    \Upon{receipt of vote request $req$ from $\mathbi{c}_f$} 
    { 
        creates a vote response $res$\;
        \If{$req.totalBlocks\geq f.totalBlocks$ and (($req.currentTermId = f.termId$ and $f.supportedId=null$) or ($req.currentTermId > f.termId$))}
        {
           $res.supported = true$ \tcp*{supports $\mathbi{c}_f$ as coordinator}   
           \tcp{records current vote option}
           $f.termId = req.currentTermId$\;
           $f.supporteId = req.candidateId$\;
        }
        \Else{
            $res.supported = false$ \tcp*{rejects $\mathbi{c}_f$}
        }
        $res.currentTermId=termId$  \tcp*{shares term ID}
        responds to candidator $\mathbi{c}_f$
     }
\end{algorithm}

\newpage
\section{Experimental Results}
\label{Appendix:experimentResults}

\setcounter{table}{0}
\renewcommand{\thetable}{T\arabic{table}}

\setlength{\textfloatsep}{4pt}
\begin{table}[!htb]
    \centering
    \renewcommand{\arraystretch}{1.3}
    \tabcaption{Performance Comparison in Time Consumption (Seconds)}
    \vspace{-1em}
    \label{tab:Appendix_results_time}
    \centering
    \scalebox{0.85}{\begin{tabular}{c|c|c|c|c|c|c}
    \hline
    \hline
    \multirow{2}{*}{\textbf{Parameter}} & \multirow{2}{*}{\textbf{Value}} & \multicolumn{5}{c}{\textbf{Schemes}} \\
    \cline{3-7}
     &  & \textbf{DataSync} & \textbf{Gossip} & \textbf{Raft} & \textbf{EDD-A} & \textbf{EdgeDis}\\
    \hline
\multirow{5}{*}{$n$}  &  8  &  526.1  &  582.8  &  566.1  &  582.1  &  290.2\\
  &  16  &  537.3  &  623.1 &  580.2  &  622.3  &  151.7\\
  &  32  &  562.9  &  689.7  &  592.5  &  688.8  &  80.8\\
  &  64  &  624.4  &  784.3  &  648.8  &  783.3  &  48.0\\
  &  128  &  795.9  &  885.8  &  751.4  &  884.7  &  32.8\\
  \hline
\multirow{6}{*}{$r$}  &  0.0\%  &  562.9  &  689.7  &  592.5  &  688.8  &  80.8\\
  &  0.2\%  &  597.8  &  731.5  &  593.5  &  731.5  &  80.9\\
  &  0.4\%  &  630.5  &  771.5  &  594.5  &  771.5  &  81.1\\
  &  0.6\%  &  661.1  &  809.1  &  595.4  &  809.1  &  81.2\\
  &  0.8\%  &  690.0  &  844.4  &  596.4  &  844.4  &  81.3\\
  &  1.0\%  &  717.0  &  877.5  &  597.4  &  877.5  &  81.5\\
  \hline
\multirow{6}{*}{$nd$}  &  1  &  562.9  &  783.4  &  604.2  &  726.8  &  81.8\\
  &  1.2  &  562.9  &  758.6  &  597.6  &  706.1  &  81.0\\
  &  1.4  &  562.9  &  737.8  &  592.5  &  688.8  &  80.3\\
  &  1.6  &  562.9  &  721.3  &  590.5  &  675.0  &  80.1\\
  &  1.8  &  562.9  &  713.0  &  589.2  &  668.1  &  79.9\\
  &  2  &  562.9  &  712.4  &  588.5  &  667.6  &  79.8\\
  \hline
\multirow{3}{*}{$dl$}  &  [5,15]  &  562.9  &  689.7  &  592.5  &  688.8  &  80.8\\
  &  [10,25]  &  562.9  &  715.2  &  624.0  &  714.3  &  84.8\\
  &  [15,40] &  562.9  &  766.2  &  687.0  &  765.3  &  92.6\\
  \hline
\multirow{5}{*}{$bs$}  &  128KB  &  843.9  &  1002.3  &  821.9  &  1001.6  &  127.8\\
  &   256KB  &  592.5  &  684.0  &  594.3  &  683.2  &  86.8\\
  &   512KB  &  562.9  &  689.7  &  592.5  &  688.8  &  80.3\\
  &   1MB  &  500.1  &  579.1  &  536.8  &  577.1  &  70.2\\
  &   2MB  &  387.5  &  440.9  &  414.2  &  439.8  &  53.3\\
  \hline
\multirow{5}{*}{$ds$}  &  256MB  &  140.7  &  172.9  &  148.1  &  172.2  &  20.1\\
  &   512MB  &  282.2  &  344.4  &  296.3  &  344.4  &  40.2\\
  &   1GB  &  562.9  &  689.7  &  592.5  &  688.8  &  80.3\\
  &   2GB  &  1125.7  &  1378.6  &  1185.1  &  1377.6  &  160.7\\
  &   4GB  &  2251.4  &  2756.4  &  2370.2  &  2755.3  &  321.3\\
    \hline
    \hline
\end{tabular}}
\end{table}

\setlength{\textfloatsep}{4pt}
\begin{table}[!htb]
    \centering
    \renewcommand{\arraystretch}{1.3}
    \tabcaption{Performance Comparison in Cost}
    \vspace{-1em}
    \label{tab:Appendix_results_cost}
    \centering
    \scalebox{0.85}{\begin{tabular}{c|c|c|c|c|c|c}
    \hline
    \hline
    \multirow{2}{*}{\textbf{Parameter}} & \multirow{2}{*}{\textbf{Value}} & \multicolumn{5}{c}{\textbf{Schemes}} \\
    \cline{3-7}
     &  & \textbf{DataSync} & \textbf{Gossip} & \textbf{Raft} & \textbf{EDD-A} & \textbf{EdgeDis}\\
    \hline
        \multirow{5}{*}{$n$}  &  8  &  160.0  &  31.2  &  28.0  &  35.6  &  28.0\\
          &  16  &  320.0  &  42.4  &  36.0  &  60.7  &  36.0\\
          &  32  &  640.0  &  64.8  &  52.0  &  100.4  &  52.0\\
          &  64  &  1280.0  &  109.6  &  84.0  &  143.8  &  84.0\\
          &  128  &  2560.0  &  199.2  &  148.0  &  215.4  &  148.0\\
    \hline
        \multirow{5}{*}{$r$}  &  0.0\%  &  640.0  &  64.8  &  52.0  &  100.4  &  52.0\\
          &  0.20\%  &  641.3  &  64.9  &  52.1  &  106.6  &  52.1\\
          &  0.40\%  &  642.6  &  65.0  &  52.2  &  112.5  &  52.2\\
          &  0.60\%  &  643.8  &  65.1  &  52.3  &  117.9  &  52.3\\
          &  0.80\%  &  645.1  &  65.2  &  52.4  &  123.1  &  52.4\\
          &  1.00\%  &  646.4  &  65.3  &  52.5  &  127.9  &  52.5\\
    \hline
        \multirow{6}{*}{$nd$}  &  1  &  640.0  &  52.0  &  52.0  &  126.1  &  52.0\\
          &  1.2  &  640.0  &  56.0  &  52.0  &  111.8  &  52.0\\
          &  1.4  &  640.0  &  60.0  &  52.0  &  100.4  &  52.0\\
          &  1.6  &  640.0  &  64.0  &  52.0  &  89.0  &  52.0\\
          &  1.8  &  640.0  &  68.0  &  52.0  &  81.9  &  52.0\\
          &  2  &  640.0  &  72.0  &  52.0  &  78.6  &  52.0\\
    \hline
       \multirow{5}{*}{$cr$} &  5  &  160.0  &  49.8  &  37.0  &  59.1  &  37.0\\
          &  10  &  320.0  &  54.8  &  42.0  &  82.0  &  42.0\\
          &  20  &  640.0  &  64.8  &  52.0  &  127.9  &  52.0\\
          &  30  &  960.0  &  74.8  &  62.0  &  173.8  &  62.0\\
          &  40  &  1280.0  &  84.8  &  72.0  &  219.6  &  72.0\\

    \hline
    \hline
\end{tabular}}
\end{table}

\end{appendices}

\end{document}